\documentclass[aps,prd,preprintnumbers,groupedaddress,nofootinbib,amssymb,eqsecnum,notitlepage]{revtex4}
\usepackage{here}
\usepackage{graphicx}
\usepackage{amsmath,amsthm,amssymb}
\usepackage{bm}

\usepackage{amsfonts}
\usepackage{dcolumn}
\usepackage{hyperref}
\allowdisplaybreaks[1]

\begin{document}
\newcommand{\newc}{\newcommand}

\newcommand{\ben}{\begin{eqnarray}}
\newcommand{\een}{\end{eqnarray}}
\newc{\be}{\begin{equation}}
\newc{\ee}{\end{equation}}
\newc{\ba}{\begin{eqnarray}}
\newc{\ea}{\end{eqnarray}}
\newc{\bea}{\begin{eqnarray*}}
\newc{\eea}{\end{eqnarray*}}
\newc{\D}{\partial}
\newc{\ie}{{\it i.e.} }
\newc{\eg}{{\it e.g.} }
\newc{\etc}{{\it etc.} }
\newc{\etal}{{\it et al.}}
\newcommand{\nn}{\nonumber}
\newc{\ra}{\rightarrow}
\newc{\lra}{\leftrightarrow}
\newc{\lsim}{\buildrel{<}\over{\sim}}
\newc{\gsim}{\buildrel{>}\over{\sim}}
\newc{\aP}{\alpha_{\rm P}}

\title{Suppression of matter couplings with a vector field 
in generalized Proca theories}

\author{
Shintaro Nakamura, 
Ryotaro Kase, and
Shinji Tsujikawa}

\affiliation{
Department of Physics, Faculty of Science, 
Tokyo University of Science, 1-3, Kagurazaka,
Shinjuku-ku, Tokyo 162-8601, Japan}

\begin{abstract}

In the context of generalized Proca theories, 
we derive the profile of a vector field $A_{\mu}$ whose 
squared $A_{\mu}A^{\mu}$ is coupled to the trace $T$ 
of matter on a static and spherically symmetric background.
The cubic Galileon self-interaction leads to the suppression of 
a longitudinal vector component due to the operation of the Vainshtein mechanism. 
For quartic and sixth-order derivative interactions, the solutions consistent with those in the continuous limit of small derivative 
couplings correspond to the branch with the 
vanishing longitudinal mode.
We compute the corrections to gravitational 
potentials outside a compact body induced by the 
vector field in the presence of cubic, quartic, and 
sixth-order derivative couplings, and show 
that the models can be consistent with local gravity constraints 
under mild bounds on the temporal vector component.
The quintic vector Galileon does not allow
regular solutions of the longitudinal mode for a rapidly decreasing matter density outside the body.

\end{abstract}

\date{\today}


\maketitle

\section{Introduction}
\label{introsec}

The problems of dark energy and dark matter imply that there 
may be additional dynamical degrees of freedom (DOFs) 
beyond those appearing in the Standard Model of 
particle physics \cite{darkreview}. 
The most studied case is a scalar field $\phi$ with a 
potential energy $V(\phi)$ \cite{quin}. Such a new scalar DOF can 
be responsible for the late-time cosmic acceleration or 
can mimic the dark matter property, depending on forms 
of the potential. There are also other scalar-field models 
with derivative self interactions---like Galileons \cite{Galileon1,Galileon2} 
and their extensions \cite{Horndeski,redis,GLPV}. 
These derivative interactions can be
the source of dark energy \cite{GSami,DT10,KTD16}
while suppressing the propagation 
of fifth forces in the Solar System  \cite{Burrage,DKT,KNT,KKY,Kase13,CKNT,APSS,KTD16va,BCLM} 
through the operation of the Vainshtein mechanism \cite{Vain}.

The scalar field is not the only possibility for explaining the dark sector 
of the Universe; the vector field can also play a similar role \cite{earlydark}. 
The standard massless Maxwell field, which respects the $U(1)$  
gauge symmetry, has two transverse electric and magnetic polarizations. 
The gauge symmetry is explicitly broken by   
introducing a mass term or derivative interactions 
of the vector field, 
in which case the longitudinal propagation emerges. 
In Refs.~\cite{Heisenberg,Tasinato,Allys,Jimenez2016}, 
the four-dimensional action of a massive Proca field
with nonminimal derivative couplings to gravity was constructed from 
the requirement of keeping two transverse and longitudinal modes 
besides two tensor polarizations arising from the gravity 
sector (see Refs.~\cite{Barrow,Durrer,TKK,Mukoh,Fleury,Hull,Lagos,Emami} 
for related works). 
In such generalized Proca theories, the equations of motion remain 
of second order, so there are no Ostrogradski instabilities. 
It is also possible to go beyond the second-order domain without 
increasing the number of propagating DOFs \cite{HKT,Kimura}.

If we apply generalized Proca theories and their extension to the isotropic 
and homogenous cosmological background, there exists an interesting 
de Sitter attractor with a constant temporal vector component \cite{cosmo}. 
The spatial vector components can be treated as perturbations 
on such a background, which is consistent with the analysis on 
the anisotropic cosmological background with time-dependent 
spatial components \cite{HKT16}. 
There are dark energy models in which all the stability conditions of 
perturbations can be consistently satisfied \cite{cosmo}. 
Moreover, the presence of an intrinsic vector mode offers the 
possibility for realizing the 
gravitational interaction weaker than that in 
general relativity (GR) \cite{Geff,NKT,DHT17}. 
This allows one to distinguish dark energy models in generalized 
Proca theories from those in GR. 

On a static and spherically symmetric background, the existence 
of hairy black hole solutions was recently studied in the context 
of generalized Proca 
theories \cite{Chagoya,Fan,Minami,Cisterna,Chagoya2,Babichev17,HKMT,HKMT2,Chagoya3}. 
For massless or massive vector fields 
without derivative interactions the spatial vector components vanish 
identically \cite{Beken72}, so the background geometry is simply described by 
the Reissner-Nordstr\"{o}m or the Schwarzschild space-time.
The existence of derivative interactions gives rise to a variety of 
hairy black hole solutions with nonvanishing temporal and longitudinal 
vector components \cite{HKMT,HKMT2}. 
This leads to the difference between two metric 
components around the black hole horizon.
The deviation from GR in the nonlinear regime of gravity can be potentially 
probed by future measurements of gravitational waves 
around black holes.

Inside the Solar System, the fifth force mediated by the vector field 
$A_{\mu}$ nonminimally coupled to gravity should be screened
for the consistency with local gravity experiments \cite{Will}. 
In Ref.~\cite{screening}, the propagation of fifth forces around a  
spherically symmetric and static body was studied in the presence of 
cubic and quartic vector Galileon interactions
under the approximation of weak gravity.
The cubic derivative interaction leads to a suppression of 
the longitudinal component $A_1$ thanks to the 
Vainshtein mechanism. The quartic derivative coupling gives rise 
to the branch $A_1=0$, so the gravitational potentials are subject  
to modifications only through the temporal vector component $A_0$. 
For both cubic and quartic interactions, the models can be compatible 
with local gravity constraints under mild bounds on $A_0$. 
This property also persists even in beyond-generalized 
Proca theories \cite{LKT16}.

The analysis of Ref.~\cite{screening} assumed that 
a direct coupling between the vector field and matter is absent, 
so the vector-matter interaction arises 
indirectly from nonminimal gravitational couplings with the vector field. 
It is not yet clear whether the existence of direct vector-matter interactions
leads to the screening of fifth forces at the level of being compatible 
with local tests of gravity.
In this paper, we will address the issue of the screening mechanism 
in the presence of a matter coupling of the form 
$QA_{\mu}A^{\mu}\,T$, where $Q$ is the coupling strength 
and $T$ is the trace of the energy-momentum tensor of matter. 
The quantum corrections to the generalized Proca action can be generated by 
matter loops, but the computation of one-loop corrections to the vector-field 
propagator shows that the theory remains 
healthy without the appearance of new ghosty DOFs  
in the domain of the effective theory \cite{Amado}. 
We will derive the vector-field profile around a compact body 
in the presence of cubic, quartic, quintic, and sixth-order-derivative 
generalized Proca interactions and estimate the corrections to 
leading-order gravitational potentials in GR.

This paper is organized as follows: 
In Sec.~\ref{eomsec} we present the full equations of motion on the 
static and spherically symmetric background in generalized Proca theories 
with the matter coupling. 
In Sec.~\ref{cubsec} we derive the profiles of temporal and 
longitudinal vector components with cubic derivative interactions
both inside and outside the body. 
We compute the corrections to gravitational potentials induced by 
the vector field and show how the Vainshtein mechanism is efficient to suppress
the propagation of fifth forces even with matter couplings.
In Secs.~\ref{quasec}, \ref{quinsec} and \ref{sixthsec} we study the propagation 
of the vector field in the presence of quartic, quintic, and sixth-order 
derivative interactions, respectively. 
While the model with quintic derivative coupling does not possess regular solutions 
of the longitudinal mode for a rapidly decreasing density profile outside the body, 
the quartic and sixth-order interactions give rise to solutions with the vanishing 
longitudinal mode. The latter models can be consistent with local gravity constraints 
under mild bounds on the temporal vector component.
Sec.~\ref{consec} is devoted to conclusions.

\section{Equations of motion}
\label{eomsec}

We consider a vector field $A_{\mu}$ with the field strength 
$F_{\mu \nu}=\nabla_{\mu}A_{\nu}-\nabla_{\nu}A_{\mu}$, 
where $\nabla_{\mu}$ is the covariant derivative operator. 
We introduce a matter perfect fluid given by 
the energy-momentum tensor
\be
T_{\mu\nu} = - \frac{2}{\sqrt{-g}} \frac{\delta (\sqrt{-g} \mathcal{L}_{m})}{\delta g^{\mu\nu}}\,,
\ee
where $g$ is a determinant of the metric tensor $g_{\mu\nu}$, 
and ${\cal L}_m$ is the matter Lagrangian density.
We assume that the vector field is coupled to  
the matter sector with the interacting 
Lagrangian density
\be
{\cal L}_{\rm coupling}=\frac{Q}{M_{\rm pl}^2}XT\,,
\label{Lcoupling}
\ee
where $Q$ is a dimensionless coupling constant, 
$M_{\rm pl}$ is the reduced Planck mass, $T$ is 
the trace of $T_{\mu \nu}$, and 
\be
X=-\frac{1}{2} A_{\mu} A^{\mu}\,.
\ee
We assume that $|Q|$ is at most of the order 1.

The generalized Proca theories are the second-order 
vector-tensor theories given by 
the action \cite{Heisenberg,Tasinato,Allys,Jimenez2016}
\be
S=\int d^{4}x \sqrt{-g} 
\left( F+\sum_{i=2}^{6}{\cal L}_i +\mathcal{L}_{m}
+\mathcal{L}_{\rm coupling}\right)\,,
\label{action}
\ee
with $F=-F_{\mu \nu} F^{\mu \nu}/4$, and
\ba
\mathcal{L}_{2}&=& G_{2}(X)-2g_4(X) F\,,
\\
\mathcal{L}_{3}&=& G_{3}(X) \nabla_{\mu} A^{\mu},
\\
\mathcal{L}_{4}&=& G_{4}(X) R 
+ G_{4,X}(X)\left[ (\nabla_{\mu} A^{\mu})^{2} 
-  \nabla_{\mu} A_{\nu} \nabla^{\nu} A^{\mu}\right]\,,
\\
\mathcal{L}_{5}&=& G_{5}(X) G_{\mu\nu} \nabla^{\mu} A^{\nu} 
- \frac{1}{6} G_{5,X} (X) \left[ (\nabla_{\mu} A^{\mu})^{3} 
- 3 \nabla_{\mu} A^{\mu} \nabla_{\rho} A_{\sigma} \nabla^{\sigma} A^{\rho} 
+ 2 \nabla_{\rho} A_{\sigma} \nabla^{\nu} A^{\rho} \nabla^{\sigma} A_{\nu} \right]
\nn
\\
&&-g_{5}(X) \tilde{F}^{\alpha\mu}\tilde{F}^{\beta}_{~\mu} \nabla_{\alpha} A_{\beta},
\\
\mathcal{L}_{6}&=& G_{6}(X) L^{\mu\nu\alpha\beta} \nabla_{\mu} A_{\nu} \nabla_{\alpha} A_{\beta}
+\frac{1}{2} G_{6,X}(X) \tilde{F}^{\alpha\beta} \tilde{F}^{\mu\nu} \nabla_{\alpha} A_{\mu} \nabla_{\beta} A_{\nu}\,,
\ea
where $G_{2,3,4,5,6}$ and $g_{4,5}$ are functions of $X$ 
with the notation $G_{i,X} \equiv \partial G_i/\partial X$, 
$R$ is the Ricci scalar, and $G_{\mu\nu}$ is the Einstein 
tensor. The tensors $\tilde{F}^{\mu\nu}$ and $L^{\mu\nu\alpha\beta}$ are defined, respectively, by 
\be
\tilde{F}^{\mu\nu} \equiv  \frac{1}{2} \mathcal{E}^{\mu\nu\alpha\beta} F_{\alpha\beta}\,,\qquad
L^{\mu\nu\alpha\beta} \equiv \frac{1}{4} \mathcal{E}^{\mu\nu\rho\sigma} \mathcal{E}^{\alpha\beta\gamma\delta} R_{\rho\sigma\gamma\delta},
\ee
where $\mathcal{E}^{\mu\nu\rho\sigma}$ is the Levi-Civit\`{a} tensor normalized by $\mathcal{E}^{\mu\nu\rho\sigma}\mathcal{E}_{\mu\nu\rho\sigma}=-4!$, and 
$R_{\rho\sigma\gamma\delta}$ is the Riemann tensor.
The theory with constant $G_6(X)$ corresponds to the 
$U(1)$ gauge-invariant derivative interaction advocated by 
Horndeski \cite{Horndeski76}. 
The Lagrangians containing the functions $g_4(X)$ and 
$g_5(X)$ correspond to intrinsic vector modes.

We consider the static and spherically symmetric background described by the line element 
\be
ds^{2} = 
-f(r) dt^{2} + h^{-1}(r)dr^{2}
+r^{2} (d\theta^{2}+\sin^{2}\theta\,d\varphi^{2}),
\label{metric}
\ee
where $f(r)$ and $h(r)$ are arbitrary functions of the distance
$r$ from the center of symmetry. 
The vector-field profile compatible with the above 
background is given by \cite{screening}
\be
A_{\mu}=\left( A_0(r), A_1(r), 0, 0 \right)\,.
\ee
The quantity $X$ can be expressed as 
$X=X_{0}+X_{1}$, where 
\be
X_0=\frac{A_0^2}{2f}\,,\qquad 
X_1=-\frac{hA_1^2}{2}\,.
\ee

For the matter sector, we consider the perfect fluid 
with the energy-momentum tensor 
$T^{\mu}_{\nu}=(\rho_m+P_m) u^{\mu}u_{\nu}
+P_m \delta^{\mu}_{\nu}$, where $\rho_m$ is the density, 
$P_m$ is the pressure,  and 
$u^{\mu}=(f^{-1/2},0,0,0)$ is the fluid four-velocity 
in the rest frame. Since the trace $T$ is given by 
\be
T=T^{\mu}_{\mu}=-\rho_m+3P_m\,,
\ee
the  interacting Lagrangian density (\ref{Lcoupling}) 
does not vanish except for the radiation ($\rho_m=3P_m$).

Varying the action (\ref{action}) with respect to $A_0$ and $A_1$, 
we obtain the equations of motion for the temporal and longitudinal 
vector components, respectively, as 
\ba
\hspace{-0.4cm}
& & 
rf \left[ 2fh(rA_0''+2A_0')+r(fh'-f'h)A_0' \right] (1-2g_4) - 2 r^2 f^2 A_0 G_{2,X}
-rfA_0 \left[ 2 rfhA_1'+(rf'h+rfh'+4fh)A_1 \right] G_{3,X}
\notag\\
\hspace{-0.4cm}
& & 
+4 f^2A_0 (rh'+h-1) G_{4,X}
-8 fA_0 \left[ rfh^2 A_1A_1'-(rf'h+rfh'+fh) X_1\right] G_{4,XX}
\notag\\
\hspace{-0.4cm}
& & 
+2r^2 fA_0' \left( 2fh^2 A_1A_1'+2f'hX_0
-2fh'X_1-hA_0A_0' \right) g_{4,X}
\notag\\
\hspace{-0.4cm}
& & 
-fA_0 \left[ f(3h-1)h'A_1+h(h-1) (f'A_1+2fA_1')   \right] G_{5,X}
-2 fhA_0X_1\left[2 fhA_1'+(f'h+fh')A_1 \right] G_{5,XX}
\notag\\
\hspace{-0.4cm}
& & 
-2 f \left[ f (3 h-1) h'A_0'+h (h-1) (2fA_0''-f'A_0') \right] G_6
-4 fh A_0'X_1 \left( h A_0 A_0'-2 fh^2 A_1 A_1'
-2 f'hX_0+2 fh'X_1 \right) G_{6,XX}
\notag\\
\hspace{-0.4cm}
& & 
-2 f \left[ 4 fh^2 X_1 A_0''-2 h (hX- X_0) f'A_0'
+2f (6h-1)h' X_1A_0'+h(h-1)A_0A_0'^2
-2 fh^2(3h-1) A_0'A_1A_1'\right] G_{6,X}
\notag\\
\hspace{-0.4cm}
& & 
-4 fh \left[ 2 rfh A_1 A_0''- \{(rf' h-3 rfh'-2fh) A_1-2 rfhA_1'\} A_0'  \right] g_5
\notag\\
\hspace{-0.4cm}
& & 
-4 rfh A_0' \left[ hA_0A_0'A_1+4 fhX_1A_1'-2 A_1(f'hX_0-fh'X_1)\right] g_{5,X}
=-\frac{2Q r^2 f^2 A_0}{M_{\rm pl}^2} 
\left( \rho_m-3P_m \right)\,,
\label{be4}
\\
\hspace{-0.4cm}
& & 
A_1 \biggl[ r^2fG_{2,X}-2 (rf'h+fh-f) G_{4,X}
+4h(rA_0 A_0'-rf' X-fX_1) G_{4,XX}-r^2hA_0'^2g_{4,X}
\nonumber \\
\hspace{-0.4cm}
& &-hA_0'^2(3h-1) G_{6,X}-2h^2X_1 A_0'^2 G_{{6,{XX}}}-\frac{fQr^2}{M_{\rm pl}^2} 
\left( \rho_m-3P_m \right) \biggr]
=r [r(f' X-A_0 A_0')+4 fX_1] G_{3,X} \nonumber \\
\hspace{-0.4cm}
& &+2 f'hX_1G_{5,X} +(A_0 A_0'-f' X)\left[ (1-h)G_{5,X}-2 hX_1G_{5,XX}\right]
-2rh A_0'^2( g_{5} +2 X_1 g_{5,X})\,,
\label{be5}
\ea
where a prime represents a derivative with respect to $r$.
To derive the gravitational equations of motion, we write the 
metric (\ref{metric}) in the form 
$ds^{2}=-f(r) dt^{2} + h^{-1}(r)dr^{2}
+r^{2}e^{2\zeta(r)} (d\theta^{2}+\sin^{2}\theta\,d\varphi^{2})$. Varying the action (\ref{action}) with respect to 
$f,h,\zeta$ and setting $\zeta=0$ at the end, 
it follows that 
\ba
\hspace{-1cm}
&& \left( c_1+\frac {c_2}{r}+\frac {c_3}{r^2}\right) h'
+c_4+\frac {c_5}{r}+\frac {c_6}{r^2}
={\cal A}_1 \,,
\label{be1}\\
\hspace{-1cm}
&&-\frac{h}{f} \left( c_1+\frac {c_2}{r}+\frac {c_3}{r^2}\right) f'
+c_7+\frac {c_8}{r}+\frac {c_9}{r^2}
= {\cal A}_2\,,
\label{be2}\\
\hspace{-1cm}
&& \left( c_{10}+\frac {c_{11}}{r} \right) f''
+ \left( c_{12}+\frac {c_{13}}{r} \right) f'^2
+ \left(\frac {c_2}{2f}+\frac {c_{14}}{r}\right) f'h'
+ \left( c_{15}+\frac {c_{16}}{r} \right) f'
+ \left(-\frac {c_8}{2h}+\frac {c_{17}}{r} \right) h'
+c_{18}+\frac {c_{19}}{r}
={\cal A}_3\,,
\label{be3}
\ea
where $c_{i}$'s $(i=1,2,\dots,19)$ are given in the Appendix, and 
\be
{\cal A}_1=\rho_m+\frac{Q}{M_{\rm pl}^2}
T \left(X_0-X_1 \right)\,,\qquad
{\cal A}_2=P_m+\frac{Q}{M_{\rm pl}^2}
T \left(X_0-X_1 \right)\,,\qquad
{\cal A}_3=-2P_m-\frac{2Q}{M_{\rm pl}^2}
T \left(X_0+X_1 \right)\,.
\label{A123}
\ee
The continuity equation of the matter sector
is given by 
\be
P_m'+\frac{f'}{2f} \left( \rho_m+P_m \right)
-\frac{Q}{M_{\rm pl}^2} \left( \rho_m' 
-3P_m' \right)X=0\,,
\label{mattereq}
\ee
which follows from Eqs.~(\ref{be4})--(\ref{be3}).

In the whole analysis of this paper, we take into account 
the Einstein-Hilbert term, such that 
\be
G_4(X) \supset \frac{M_{\rm pl}^2}{2}\,,
\label{Hilbert}
\ee
where $M_{\rm pl}$ is the reduced Planck mass. 
If there exist derivative couplings $G_i(X)$ and $g_i(X)$ with even indices $i$ alone,  Eq.~(\ref{be5}) reduces to 
\be
A_1 \left[ {\cal F} (A_1,A_0,A_0', f, h, f') 
-\frac{fQr^2}{M_{\rm pl}^2} \left( 
\rho_m-3P_m \right) \right]=0\,.
\label{A1re}
\ee
The function ${\cal F}$ consists of the sum of 
terms written in the forms $\beta_i \tilde{{\cal F}}_i$ 
(with $i=2,4,6$) and $\gamma_4 \tilde{{\cal G}}_4$, where 
$\beta_i$ and $\gamma_4$ are coupling constants in 
$G_i(X)$ and $g_4(X)$, respectively, and 
$\tilde{{\cal F}}_i$ and $\tilde{{\cal G}}_4$ are regular functions of $A_0, A_1, f, h$ and their derivatives. 
{}From Eq.~(\ref{A1re}), the vector component 
$A_1$ has two branches satisfying (i) $A_1=0$, 
or (ii) ${\cal F}=fQ r^2 (\rho_m-3P_m)/M_{\rm pl}^2$. 
In the limit that $\beta_i \to 0$ and $\gamma_4 \to 0$, 
the function ${\cal F}$ vanishes, so branch (ii) 
does not exist for nonrelativistic matter ($P_m \ll \rho_m$) 
with a nonvanishing coupling constant $Q$.
Hence, the branch consistent with the general-relativistic 
limit corresponds to $A_1=0$.
This property does not generally hold for the couplings  
$G_i(X)$ and $g_i(X)$ with odd indices $i$, in which case 
the nonvanishing branch of $A_1$ can arise 
from Eq.~(\ref{be5}).

We have derived the field equations for general 
theories given by the action (\ref{action}), but 
we will focus on the theories with 
\be
g_4(X)=0\,,\qquad g_5(X)=0\,,
\ee
in the following discussion. 
This reflects the fact that the analysis with the derivative couplings $G_{3,4,5,6}(X)$ is 
sufficiently general to understand basic properties of 
the screening mechanism in generalized Proca theories.

As we will see in subsequent sections, the temporal vector 
component $A_0$ is generally dominated by a constant 
$a_0$ with a small variation around it. 
In such cases, it is convenient to express the quantity 
$X_0$ in the form
\be
X_0=\tilde{X}_0+\delta X_0\,,\qquad
\tilde{X}_0=\frac{a_0^2}{2f(r)}\,,
\label{X0a0}
\ee
where $\delta X_0$ characterizes the deviation from 
$\tilde{X}_0$. 
Unless $a_0^2$ is very much smaller than $M_{\rm pl}^2$, 
the last term on the lhs of Eq.~(\ref{mattereq}) gives rise to 
a large modification to the pressure $P_m$ relative
to the $Q=0$ case.
By defining 
\ba
\rho &\equiv& 
\rho_m-\frac{Q}{M_{\rm pl}^2} \left( \rho_m-3P_m 
\right)\tilde{X}_0\,,
\label{rhodef}\\
P &\equiv& 
P_m-\frac{Q}{M_{\rm pl}^2} \left( \rho_m-3P_m 
\right)\tilde{X}_0\,,
\label{Ppre}
\ea
it is possible to rewrite Eq.~(\ref{mattereq}) 
without having the contribution from $\tilde{X}_0$.
Expressing $\rho_m$ and $P_m$ in terms of $\rho$ and $P$
and substituting them into Eq.~(\ref{mattereq}), 
it follows that 
\be
P'+\frac{f'}{2f} \left( \rho+P \right)
+\frac{Q[M_{\rm pl}^2 f^2 T_*'
+Qa_0^2 (fT_*)'](\delta X_0+X_1)}{(M_{\rm pl}^2 f+Qa_0^2)^2}=0\,,
\label{mattereq2}
\ee
where
\be
T_* \equiv -\rho+3P=\left(1+\frac{Qa_0^2}{M_{\rm pl}^2 f} \right)T\,.
\label{TsT}
\ee
For nonrelativistic matter characterized by $P_m \ll \rho_m$, 
the matter-coupling term in Eq.~(\ref{Ppre}) does not dominate over 
$P_m$ under the condition 
\be
\left| \frac{Q}{M_{\rm pl}^2} \rho_m \tilde{X}_0 \right|
\lesssim P_m\,,
\label{Qconas}
\ee
which is assumed in the following discussion.
As long as the screening mechanism operates to suppress 
both $X_1$ and $\delta X_0$, the $Q$-dependent terms 
in Eq.~(\ref{mattereq2}) work as tiny corrections to the 
continuity equation $P'+f'(\rho+P)/(2f)=0$ in GR. 
We also note that the terms on the rhs of 
Eqs.~(\ref{be1})--(\ref{be3}) are expressed, 
respectively, as
\be
{\cal A}_1=\rho+\frac{\tilde{Q}}{M_{\rm pl}^2} T_*
 (\delta X_0-X_1)\,,\qquad
{\cal A}_2=P+\frac{\tilde{Q}}{M_{\rm pl}^2} T_*
 (\delta X_0-X_1)\,,\qquad
{\cal A}_3=-2P-\frac{2\tilde{Q}}{M_{\rm pl}^2} 
T_* (\delta X_0+X_1)\,,
\label{A123d}
\ee
where 
\be
\tilde{Q} \equiv \frac{Q}{1+Qa_0^2/(M_{\rm pl}^2 f)}\,.
\label{tQ}
\ee
Unlike Eq.~(\ref{A123}), the matter-coupling terms in 
Eq.~(\ref{A123d}) do not contain $\tilde{X}_0$.
For nonrelativistic matter, the condition 
(\ref{Qconas}) translates to $|Qa_0^2| \ll M_{\rm pl}^2 f$. 
Then, the coupling $\tilde{Q}$ is approximately equivalent to 
$Q$ with $T_* \simeq T$ in Eq.~(\ref{TsT}).

We will exploit the rescaled energy density 
$\rho$ and the pressure $P$ to discuss the profile of 
a spherically symmetric body with radius $r_*$. 
To derive solutions to the vector-field and
gravitational potentials, we consider the following 
matter density profile:
\ba
\rho(r) = \left\{ \begin{array}{ll}
\rho_0 & \quad ({\rm for}~~r<r_*) \\
\rho_0 \mu & \quad ({\rm for}~~r>r_*)
\end{array} \right. \,,
\label{mudef}
\ea
where $\rho_0$ is a constant density, and $\mu$ is 
a dimensionless constant much smaller than 1. 
In Sec.~\ref{cubsec} we will also numerically obtain 
solutions to the vector field for a varying matter density. 
The Schwarzschild radius of the source is defined by 
\be
r_g(r) \equiv \frac{1}{M_{\rm pl}^2} \int_0^r 
\rho (\tilde{r}) \tilde{r}^2 d\tilde{r}\,.
\ee
Taking the vacuum limit $\mu \to 0$ outside the body, 
it follows that $r_g \simeq \rho_0r_*^3/(3M_{\rm pl}^2)$. 
We introduce the gravitational 
potentials $\Psi$ and $\Phi$, as 
\be
f(r)=e^{2\Psi(r)}\,,\qquad
h(r)=e^{-2\Phi(r)}\,,
\ee
and employ the weak gravity approximation
\be
\Phi_* \equiv \frac{\rho_0r_*^2}{M_{\rm pl}^2} \ll 1\,,
\label{weakap}
\ee
which amounts to the condition $r_g \ll r_*$.

In GR without the coupling $Q$, the gravitational potentials 
$\Psi$ and $\Phi$ inside the body ($r<r_*$) 
are given by the internal Schwarzschild solution 
\be
e^{\Psi_{\rm GR}}=
\frac32 \sqrt{1-\frac{\Phi_*}{3}}
-\frac12 \sqrt{1-\frac{\Phi_*}{3} 
\frac{r^2}{r_*^2}}\,,\qquad \quad
e^{\Phi_{\rm GR}}=
\left( 1-\frac{\Phi_*}{3} 
\frac{r^2}{r_*^2} \right)^{-1/2}\,,
\label{gra1GR}
\ee
with the pressure 
\be
P=\frac{\sqrt{1-\Phi_*r^2/(3r_*^2)}-\sqrt{1-\Phi_*/3}}
{3\sqrt{1-\Phi_*/3}-\sqrt{1-\Phi_*r^2/(3r_*^2)}}\,\rho_0\,.
\label{pressure}
\ee
Under the approximation (\ref{weakap}), 
the ratio between $P$ and $\rho_0$ can be estimated 
as $P/\rho_0 \simeq \Phi_* (1-r^2/r_*^2)/12 \ll 1$. 
In such cases, the condition (\ref{Qconas}) translates to 
\be
\left| \frac{Qa_0^2}{M_{\rm pl}^2 f} \right| 
\lesssim \Phi_*\,.
\label{Qa0}
\ee
In the presence of the vector field coupled to matter, 
the gravitational potentials (\ref{gra1GR}) are  
subject to modifications due to the existence of the 
$\tilde{Q}$-dependent terms in Eq.~(\ref{A123d}).
Provided that the two conditions 
\be
\left| \tilde{Q}\,\frac{\delta X_0}{M_{\rm pl}^2} \right| \ll \Phi_*\,,\qquad
\left| \tilde{Q}\,\frac{X_1}{M_{\rm pl}^2} \right| \ll \Phi_*
\label{Qcon}
\ee
are satisfied, the contributions arising from the matter 
coupling can be treated as corrections to 
the leading-order gravitational potentials (\ref{gra1GR}). 
Under the operation of the screening mechanism, we will 
show that it is possible to satisfy the conditions (\ref{Qcon}).

In the vacuum limit, the gravitational potentials  
outside the body in GR are given by 
\be
e^{\Psi_{\rm GR}}
=\left( 1-\frac{\Phi_*}{3}\frac{r_*}{r} \right)^{1/2}\,,
\qquad \quad
e^{\Phi_{\rm GR}}=
\left( 1-\frac{\Phi_*}{3}\frac{r_*}{r} \right)^{-1/2}\,.
\label{gra2GR}
\ee
The presence of the vector field coupled to matter gives rise 
to modifications to $\Psi_{\rm GR}$ and $\Phi_{\rm GR}$, but they can be 
again treated as corrections to the leading-order gravitational potentials 
for the vector components obeying the conditions of Eq. (\ref{Qcon}).

For the comparison with the results derived for $Q=0$ 
in Ref.~\cite{screening}, we will adopt the notations
\be
\phi(r)=-\frac{A_0(r)}{f(r)}\,,\qquad 
\chi'(r)=A_1(r)\,,
\ee
where $\chi(r)$ is a longitudinal scalar. The transverse 
vector mode in $A_1$ vanishes due to its regularity 
at $r=0$ \cite{screening}. 
By using $\rho$ and $P$ defined by Eqs.~(\ref{rhodef}) and 
(\ref{Ppre}) as well as the coupling $\tilde{Q}$ given by 
Eq.~(\ref{tQ}), the terms containing $Q$ in Eqs.~(\ref{be4}) 
and (\ref{be5}) can be expressed as 
$-2\tilde{Q} r^2f^2A_0 (\rho-3P)/M_{\rm pl}^2$ 
and $-f\tilde{Q}r^2A_1(\rho-3P)/M_{\rm pl}^2$, 
respectively.
We consider nonrelativistic matter satisfying 
$P \ll \rho$ and employ the approximation 
$\tilde{Q} \simeq Q$ under the condition (\ref{Qa0}).

\section{Cubic vector Galileons}
\label{cubsec}

Let us begin with the cubic Galileon model 
given by the functions
\be
G_2=m^2 X\,,\qquad 
G_3=\beta_3 X\,,\qquad 
G_4=\frac{M_{\rm pl}^2}{2}\,,\qquad
G_5=0\,,\qquad
G_6=0\,,\qquad
\ee
where $m$ is a constant having a dimension of mass, 
and $\beta_3$ is a dimensionless constant.
{}From Eqs.~(\ref{be4}) and (\ref{be5}), 
we obtain the following 
equations of motion: 
\ba
& &
\frac{1}{r^2} \frac{d}{dr} (r^2 \phi')
-\beta_3 \phi\, \frac{1}{r^2} \frac{d}{dr} (r^2 \chi') 
+2\phi \left( \Psi''+\Psi'^2-\Psi'\Phi' \right)
-\left( \beta_3 \phi \chi'-3\phi'-\frac{4\phi}{r} \right)
\Psi'
+\left( \beta_3 \phi \chi'-\phi' \right)\Phi'  \nonumber \\
& &
=-e^{2\Phi} \phi \left( \frac{\tilde{Q} \rho}{M_{\rm pl}^2}
-m^2 \right)\,,
\label{phieq1} \\
& &
\beta_3 \left[ e^{2\Psi} \phi \phi'+\frac{2}{r}
e^{-2\Phi} \chi'^2+(e^{2\Psi} \phi^2
+e^{-2\Phi}\chi'^2)\Psi' \right]=
\left( \frac{\tilde{Q} \rho}{M_{\rm pl}^2}
-m^2 \right) \chi'\,.
\label{chieq1}
\ea
For $|m|$ smaller than the order of
the today's Hubble expansion rate $H_0 \simeq 10^{-33}$ eV, 
the term $|\tilde{Q}\rho/M_{\rm pl}^2|$ is much larger than 
$|m^2|$ for $|\tilde{Q}|\rho \gg \rho_c$, where 
$\rho_c \simeq 10^{-29}$ g/cm$^3$ is today's 
critical density.
Unless $|Q|$ is extremely smaller than 1, the condition 
$|\tilde{Q}|\rho \gg \rho_c$ is well satisfied in the Solar System. 
Hence, we will ignore the term $m^2$ relative to 
$\tilde{Q}\rho/M_{\rm pl}^2$
in the whole analysis of this paper.
Under the condition (\ref{Qa0}), the coupling $\tilde{Q}$ 
defined by Eq.~(\ref{tQ}) is at most of the order 
$Q[1+{\cal O}(\Phi_*)]$. We neglect the 
contribution of the order  $\Phi_*$ 
in $\tilde{Q}$ for the estimations 
of $\phi$ and $\chi'$, so that the term 
$\tilde{Q} \rho/M_{\rm pl}^2$ appearing on the rhs of 
Eqs.~(\ref{phieq1}) and (\ref{chieq1}) is 
approximated as $Q \rho/M_{\rm pl}^2$.

We deal with the general-relativistic gravitational potentials 
$\Psi_{\rm GR}$ and $\Phi_{\rm GR}$ as leading-order 
contributions to $\Psi$ and $\Phi$, respectively, 
and substitute them into Eqs.~(\ref{phieq1}) and (\ref{chieq1}) 
to obtain the solutions to $\phi$ and $\chi'$. 
We then plug the solutions of vector-field profiles into
Eqs.~(\ref{be1}) and (\ref{be2}) to derive corrections 
to the leading-order gravitational potentials outside the body.

\subsection{$r<r_*$}

We first derive the field profiles for the distance $r$ smaller than $r_*$.  
Substituting Eq.~(\ref{gra1GR}) into 
Eqs.~(\ref{phieq1}) and (\ref{chieq1}),  we obtain
\ba
& &
\frac{d}{dr} \left( r^2 \phi' \right) 
-\beta_3 \phi \frac{d}{dr} \left( r^2 \chi' \right)
+(1+Q) \phi \Phi_* \frac{r^2}{r_*^2}
+\beta_3 \phi \chi' \Phi_*
\frac{r^3}{6r_*^2} \simeq 0\,,
\label{phieqG3} \\
& &
\beta_3 \left( \phi \phi'+\frac{2}{r}\chi'^2
+\frac{\phi^2 \Phi_*}{6r_*^2}r \right) 
\simeq Q \frac{\Phi_*}{r_*^2}\chi'\,.
\label{chieqG3}
\ea
{}From Eq.~(\ref{chieqG3}), it follows that 
\be
\chi'(r)=\frac{Q\Phi_*}{4\beta_3 r_*^2} 
\left[ 1 - \sqrt{1-\frac{8\beta_3^2r_*^4 \phi}
{Q^2 \Phi_*^2r} \left(\phi'+\frac{\phi \Phi_*}{6r_*^2}r
\right)}\right]r\,, 
\label{chisoG3}
\ee
where we have chosen the branch recovering 
the solution $\chi'(r) \to 0$ in the continuous limit 
$\beta_3 \to 0$. We will focus on the positive 
derivative coupling  
\be
\beta_3>0\,,
\ee
but we will not restrict the signs of $Q$. 

Analogous to the discussion given in Ref.~\cite{screening}, 
we search for the solution where the temporal vector component 
$\phi$ is close to a positive constant 
$\phi_0$, such that 
\be
\phi(r)=\phi_0+f(r)\,,\qquad |f(r)| \ll \phi_0\,.
\label{phiasu}
\ee
In what follows, we identify the constant $a_0$ 
in Eq.~(\ref{X0a0}) with $\phi_0$. 
After deriving the solutions to $\phi(r)$ and $\chi'(r)$ under 
the assumption (\ref{phiasu}), we can confirm that the 
term $\beta_3 \phi \chi' \Phi_* r^3/(6r_*^2)$ 
in Eq.~(\ref{phieqG3}) is negligible relative to 
other contributions. Integrating Eq.~(\ref{phieqG3}) 
after replacing $\phi$ with $\phi_0$, we obtain
\be
r^2 \phi'-\beta_3 \phi_0 r^2 \chi'
+(1+Q)\phi_0\Phi_* \frac{r^3}{3r_*^2}=0\,,
\label{phiint}
\ee
where the integration constant is set to 0 to satisfy 
the boundary condition $\phi'(0)=0$. 
Now, we substitute Eq.~(\ref{chisoG3}) into Eq.~(\ref{phiint}) 
by replacing $\phi$ with $\phi_0$ and then 
solve Eq.~(\ref{phiint}) for $\phi'(r)$. 
This process leads to 
\ba
\phi'(r) &=&-\frac{\phi_0 \Phi_*{\cal F}_1}{3r_*^2}r
\,,\label{phidr}\\
\phi(r) &=& \phi_0 \left( 1-\frac{\Phi_* {\cal F}_1}{6}
\frac{r^2}{r_*^2} \right)\,,
\label{phiG3}
\ea
where 
\ba
{\cal F}_1
&\equiv& 
s_{\beta_3}+1+\frac14 Q-\sqrt{s_{\beta_3}^2
+\left( 1+\frac12 Q 
\right)s_{\beta_3}+\frac{9}{16}Q^2}\,,\\
s_{\beta_3}
&\equiv& \frac{3(\beta_3 \phi_0 M_{\rm pl})^2}
{4\rho_0}=\frac{3(\beta_3 \phi_0 r_*)^2}{4\Phi_*}\,.
\ea
Substituting Eq.~(\ref{phidr}) into 
Eq.~(\ref{chisoG3}), we obtain
\be
\chi'(r)= 
\phi_0 \frac{Q}{8} \sqrt{\frac{3\Phi_*}{s_{\beta_3}}}
\left[ 1-\sqrt{1+\frac{32s_{\beta_3}}{9Q^2}
\left( {\cal F}_1-\frac{1}{2} \right)} \right]
\frac{r}{r_*}\,.
\label{chiG3}
\ee

In the limit that $s_{\beta_3} \to 0$, we have 
${\cal F}_1=1-Q/2$ for $Q>0$ and 
${\cal F}_1 \simeq 1+Q$ for $Q<0$. 
{}From Eq.~(\ref{phiG3}), the difference $|\phi(r)/\phi_0-1|$ 
is of the order $\Phi_*$ around $r=r_*$. 
Since the quantity $\delta X_0$ in Eq.~(\ref{X0a0}) is at most 
of the order $\phi_0^2\Phi_*$, the first condition of 
Eq.~(\ref{Qcon}) is satisfied for $|Q\phi_0^2/M_{\rm pl}^2| \ll 1$. 
Indeed, this latter condition holds under the requirement (\ref{Qa0}).
Taking the limit $s_{\beta_3} \to 0$, the longitudinal 
mode (\ref{chiG3}) reduces to 
\be
\chi'(r)
\simeq
\phi_0 \sqrt{\Phi_* s_{\beta_3}}\, 
{\cal G}_1 \frac{r}{r_*}\,,
\label{chiG3s}
\ee
where ${\cal G}_1=\sqrt{3} (Q-1)/(9Q)$ for $Q>0$ and 
${\cal G}_1=-\sqrt{3} (1+2Q)/(9Q)$ for $Q<0$. 
In the small-coupling limits $Q \to 0^+$ and $Q \to 0^-$, 
we have $\chi'(r)<0$ and $\chi'(r)>0$, respectively.
The amplitude $|\chi'(r)|$ 
reaches the maximum $\phi_0 \sqrt{\Phi_* s_{\beta_3}}\, 
|{\cal G}_1|$ at $r=r_*$. 
Provided that $s_{\beta_3} \lesssim Q^2$, 
the second condition of Eq.~(\ref{Qcon}) holds 
under the requirement of Eq. (\ref{Qa0}).

Taking another limit $s_{\beta_3} \to \infty$, 
it follows that ${\cal F}_1 \simeq 
1/2+(1-Q)(1+2Q)/(8s_{\beta_3})$. 
In this case, the solutions (\ref{phiG3}) 
and (\ref{chiG3}) reduce, respectively, to 
\ba
\phi(r) 
&\simeq& \phi_0 \left( 1-\frac{\Phi_*}{12}
\frac{r^2}{r_*^2} \right)\,,
\label{phiG3l}\\
\chi'(r)
&\simeq& 
\phi_0
\sqrt{\frac{\Phi_*}{s_{\beta_3}}}{\cal G}_2
\frac{r}{r_*}\,,
\label{chiG3l}
\ea
where ${\cal G}_2=\sqrt{3}(Q-1)/12$ for $Q>0$ 
and ${\cal G}_2=\sqrt{3}(3Q+|Q+2|)/24$ 
for $Q<0$.
For $s_{\beta_3} \gtrsim {\cal O}(1)$, the cubic 
Galileon coupling leads to the strong suppression 
of the longitudinal mode due to the operation of 
the Vainshtein mechanism. 
Again, the two conditions (\ref{Qcon}) are consistently 
satisfied under the requirement (\ref{Qa0}).
The value of $s_{\beta_3}$ can be estimated as
\be
s_{\beta_3} \simeq 6 \times 10^{90} 
\beta_3^2 \left( \frac{\phi_0}{M_{\rm pl}} 
\right)^2 \left(\frac{1~{\rm g/cm}^3}{\rho_0} 
\right)\,,
\ee
and hence it can be naturally larger than unity 
for density of the order $\rho_0=1~$g/cm$^3$ 
(like the Sun or the Earth).

\subsection{$r>r_*$}

Let us derive the solutions to $\phi(r)$ and $\chi'(r)$ 
outside the spherically symmetric body. 
On using the leading-order gravitational potentials 
(\ref{gra2GR}), Eqs.~(\ref{phieq1}) and (\ref{chieq1}) 
reduce, respectively, to\footnote{In Ref.~\cite{screening}, 
the approximate gravitational potentials 
$\Psi_{\rm GR} \simeq -\Phi_*r_*/(6r)$ and 
$\Phi_{\rm GR} \simeq \Phi_*r_*/(6r)$ were used 
instead of Eq.~(\ref{gra2GR}) for the derivation of 
the vector-field equations of motion. In this case, the 
extra term $\phi \Phi_* r_*^2/(9r^2)$ arises on 
the lhs of Eq.~(\ref{phieqG3d}), 
but this does not affect the discussion 
after Eq.~(\ref{philm}).} 
\ba
& &
\frac{d}{dr} \left( r^2 \phi' \right) 
-\beta_3 \phi \frac{d}{dr} \left( r^2 \chi' \right)
\simeq -\frac{Q\phi \rho_0 \mu(r)}{M_{\rm pl}^2}r^2\,,
\label{phieqG3d} \\
& &
\beta_3 \left( \phi \phi'+\frac{2}{r}\chi'^2
+\frac{\phi^2 \Phi_*r_*}{6r^2} \right) 
\simeq \frac{Q\rho_0 \mu(r) \chi'}{M_{\rm pl}^2}\,.
\label{chieqG3d}
\ea
We assume that the quantity $\mu (r)$, which is defined 
by Eq.~(\ref{mudef}), is a constant $\mu$ much 
smaller than 1.
{}From Eq.~(\ref{chieqG3d}), the longitudinal mode is
expressed as 
\be
\chi'(r)=\frac{Q\Phi_*\mu}{4\beta_3 r_*^2} 
\left[ 1-\sqrt{1-\frac{8\beta_3^2r_*^4 \phi}
{\mu^2Q^2 \Phi_*^2r} \left(\phi'+\frac{\phi \Phi_* r_*}
{6r^2}\right)}\right]r\,.
\label{chisoG32}
\ee
On using the approximation scheme (\ref{phiasu}), 
we can integrate Eq.~(\ref{phieqG3d}) to give 
\be
r^2 \phi'-\beta_3 \phi_0 r^2 \chi'+\frac{Q\phi_0 \Phi_* \mu}
{3r_*^2}r^3 \simeq {\cal C}\,,
\label{philm}
\ee
where the integration constant ${\cal C}$ is fixed to be
${\cal C}=-\phi_0 \Phi_* r_*(1+Q-Q\mu )/3$ by 
matching the solution (\ref{philm}) with 
Eq.~(\ref{phiint}) at $r=r_*$. 
Then, we obtain the following relation: 
\be
\phi'-\beta_3 \phi_0 \chi'+\frac{Q\phi_0 \Phi_* \mu}
{3r_*^2}r=-\frac{(1+Q-Q \mu)\phi_0 \Phi_* r_*}{3r^2}\,.
\label{phirela}
\ee
Substituting Eq.~(\ref{chisoG32}) into Eq.~(\ref{phirela}) 
under the approximation (\ref{phiasu}), it follows that 
\ba
\phi'(r) &=&-\frac{\phi_0 \Phi_* r_*}{3r^2}
{\cal F}_2(r)\,,\label{phidr2}
\ea
where 
\ba
\hspace{-0.3cm}
{\cal F}_2(r) 
&\equiv& 1+\xi(r)+Q(1-\mu)
+\frac{Q\mu}{4s_{\beta_3}}\xi(r)
-\sqrt{\xi(r) \left[
1+\xi(r)+2Q+\frac{Q\mu}{2s_{\beta_3}} \left( \xi(r)-4s_{\beta_3}\right)
\right]+\left(\frac{3Q\mu}{4s_{\beta_3}} \xi(r) \right)^2}\,,
\label{F2}\\
\hspace{-0.3cm}
\xi(r)
&\equiv& s_{\beta_3} \frac{r^3}{r_*^3}\,.
\ea
\subsubsection{$s_{\beta_3} \gg 1$}

We first study the case where $s_{\beta_3} \gg 1$. 
Since $\xi(r) \gg 1$ outside the body, 
we first take the limit $\xi(r) \to \infty$ in Eq.~(\ref{F2}). 
The ratio $\mu$ is much smaller than $1$, so 
we carry out the expansion of 
${\cal F}_2 (\xi(r) \to \infty)$ around $\mu=0$. 
This process leads to 
\be
{\cal F}_2(r) \simeq 
\frac12+\frac{(1+2Q)^2}{8s_{\beta_3}}\frac{r_*^3}{r^3}
+\frac{1+2Q}{8s_{\beta_3}}Q\mu+\cdots\,,
\label{F2ap}
\ee
where we use the approximation $\xi (r) \gg s_{\beta_3}$ 
(i.e., $r \gg r_*$) for deriving the third term 
on the rhs of ${\cal F}_2(r)$.
For the distance $r$ satisfying 
\be
r \ll r_V \equiv \left| \frac{1+2Q}{Q\mu} 
\right|^{1/3} r_*\,,
\label{rV}
\ee
the second term on the rhs of Eq.~(\ref{F2ap}) dominates 
over the third one. In this regime, Eqs.~(\ref{phidr2}) and  
(\ref{chisoG32}) reduce, respectively, to
\ba
\phi'(r) &\simeq&
-\frac{\phi_0 \Phi_* r_*}{3r^2} 
\left[ \frac12+\frac{(1+2Q)^2}{8s_{\beta_3}}
\frac{r_*^3}{r^3} \right]\,,
\label{phiva}\\
\chi'(r) &\simeq&
\frac{\phi_0 Q\mu}{8}
\sqrt{\frac{3\Phi_*}{s_{\beta_3}}} \frac{r}{r_*} 
\left[ 1 -\sqrt{1+\frac{4(1+2Q)^2}{9Q^2\mu^2}
\frac{r_*^6}{r^6}} 
\right]\,.
\label{chiva}
\ea
Since the second term in the square bracket 
of Eq.~(\ref{phiva}) rapidly approaches 0 
for increasing $r$, we obtain 
the approximate integrated solution
\be
\phi(r) \simeq 
\phi_0 \left(1+\frac{\Phi_*}{6}
\frac{r_*}{r}-\frac{\Phi_*}{4} \right)\,,
\label{phival}
\ee
where we have performed the matching of the solution with 
Eq.~(\ref{phiG3l}) at $r=r_*$.
This shows that $\phi(r)$ is nearly constant around $\phi_0$.
Under the condition (\ref{rV}), the second term in the square root of Eq.~(\ref{chiva}) is much larger than 1, so the longitudinal mode reduces to 
\be
\chi'(r) \simeq  -\frac{\phi_0}{12}
\sqrt{\frac{3\Phi_*}{s_{\beta_3}}}\, 
\eta_1 \frac{r_*^2}{r^2}\,,\qquad \quad
\eta_1 \equiv \frac{Q}{|Q|} |1+2Q| \,.
\label{chiva2}
\ee
{}From Eqs.~(\ref{phival}) and (\ref{chiva2}), we find that 
the two conditions of Eq. (\ref{Qcon}) are 
satisfied under the requirement (\ref{Qa0}).

The distance $r_V$ can be regarded as the Vainshtein radius, 
within which the propagation of the longitudinal mode is suppressed due to the existence of 
cubic Galileon interactions. 
For $|Q|$ of the order of unity, the Vainshtein radius can be 
estimated as $r_V \simeq \mu^{-1/3}r_*$. 
The density $\rho_0$ is related to the Schwarzschild radius 
$r_g$, as $\rho_0 \simeq 3M_{\rm pl}^2r_g/r_*^3$.
If $\mu \rho_0$ is of the order of the present 
cosmological density, 
it is related to today's Hubble radius, 
$r_H \simeq 10^{28}$ cm, as 
$\mu \rho_0 \simeq 3M_{\rm pl}^2/r_H^2$.
Then, the Vainshtein radius is of the order of
$r_V \simeq (r_g r_H^2)^{1/3}$. 
For the Sun ($r_g \simeq 10^5$ cm), we have 
$r_V \simeq 10^{20}$ cm, which is much larger 
than the Solar-System scale.
Thus, the propagation of the longitudinal mode is 
suppressed inside the Solar System thanks to 
the Vainshtein mechanism.

\subsubsection{$s_{\beta_3} \ll 1$}

We proceed to the case in which $s_{\beta_3}$ 
is much smaller than 1.
We introduce the critical distance 
\be
r_c=\frac{r_*}{s_{\beta_3}^{1/3}}\,.
\label{rc}
\ee
For the distance $r_*<r \ll r_c$, the quantity $\xi(r)$ 
is much smaller than 1.
Expanding Eq.~(\ref{F2}) around $\xi(r)=0$, it follows that 
${\cal F}_2(r) \simeq 1+Q(1-\mu)-\sqrt{[1+2Q(1-\mu)]\xi(r)}$. 
Ignoring the contribution of the $\xi(r)$-dependent term 
in ${\cal F}_2(r)$, Eqs.~(\ref{phidr2}) and  
(\ref{chisoG32}) reduce, respectively, to
\ba
\phi'(r) 
&\simeq& -\frac{\phi_0 \Phi_* r_*}{3r^2}
\left[ 1+Q(1-\mu) \right]\,,
\label{phiG3sd}\\
\chi'(r)
&\simeq& \frac{\phi_0Q\mu}{8} \sqrt{\frac{3\Phi_*}{s_{\beta_3}}} 
\frac{r}{r_*} \left[ 1-\sqrt{1+
\frac{16[1+2Q(1-\mu)]}{9(Q\mu)^2} 
s_{\beta_3}\frac{r_*^3}{r^3}}\right]\,,
\label{chiG3sd}
\ea
so that the temporal component stays nearly constant 
around $\phi_0$.

If the coupling $s_{\beta_3}$ satisfies the condition 
\be
s_{\beta_3} \ll (Q\mu)^2\,,
\ee
the magnitude of the second term in the square 
root of Eq.~(\ref{chiG3sd}) is much smaller than 1. 
In this case, Eq.~(\ref{chiG3sd}) yields 
\be
\chi'(r) \simeq -\frac{\phi_0}{9}
 [1+2Q(1-\mu)]  \sqrt{3\Phi_*}\frac{\sqrt{s_{\beta_3}}}
{Q\mu} \frac{r_*^2}{r^2}\,,
\label{chirsm}
\ee
so the radial dependence of $|\chi'(r)|$ is similar to Eq.~(\ref{chiva2}) with the suppression of the order
$\sqrt{\Phi_*}\sqrt{s_{\beta_3}}/(Q\mu)(r_*^2/r^2)$ 
relative to $\phi_0$.  In the limit that 
$\beta_3 \to 0$, the longitudinal mode vanishes.

For the intermediate coupling strength 
$s_{\beta_3}$ satisfying 
\be
|Q|\mu \ll s_{\beta_3} \ll 1\,,
\label{Qeq}
\ee
the magnitude of the second term in the square root of Eq.~(\ref{chiG3sd}) 
is much larger than 1 for 
the distance $r_*<r<r_c$. 
Then, the longitudinal mode reduces to 
\be
\chi'(r) \simeq -\frac{\phi_0}{6} \eta_2
\sqrt{3\Phi_* \frac{r_*}{r}}\,, 
\qquad \quad \eta_2 \equiv \frac{Q}{|Q|} 
\sqrt{1+2Q(1-\mu)}\,,
\label{chisr}
\ee
which decreases more slowly relative to 
Eqs.~(\ref{chiva2}) and (\ref{chirsm}).
Note that the existence of the solution (\ref{chisr}) 
requires the condition $1+2Q(1-\mu)>0$. 
The field profiles derived above satisfy the 
two consistency conditions of Eq.~(\ref{Qcon}) 
under the requirement (\ref{Qa0}).
For the coupling $s_{\beta_3}$ satisfying 
$(Q\mu)^2 \ll s_{\beta_3} \ll |Q|\mu$, 
the solution to $\chi'(r)$ is given by Eq.~(\ref{chisr}) 
for the distance $r_*<r \lesssim r_t \equiv
[s_{\beta_3}/(Q\mu)^2]^{1/3}$ and by 
Eq.~(\ref{chirsm}) for $r_t \lesssim r<r_c$.

Since $\xi(r) \gg 1$ for the distance $r_c \ll r \ll r_V$, 
the solutions to $\phi'(r)$ and $\chi'(r)$ are described 
by Eqs.~(\ref{phiva}) and (\ref{chiva2}), respectively.

\subsection{Vector-field profile for
 varying matter density}

%%%%%%%%%%%%%%%%%%%%%%%%%%%%
\begin{figure}
\begin{center}
\includegraphics[width=3.2in]{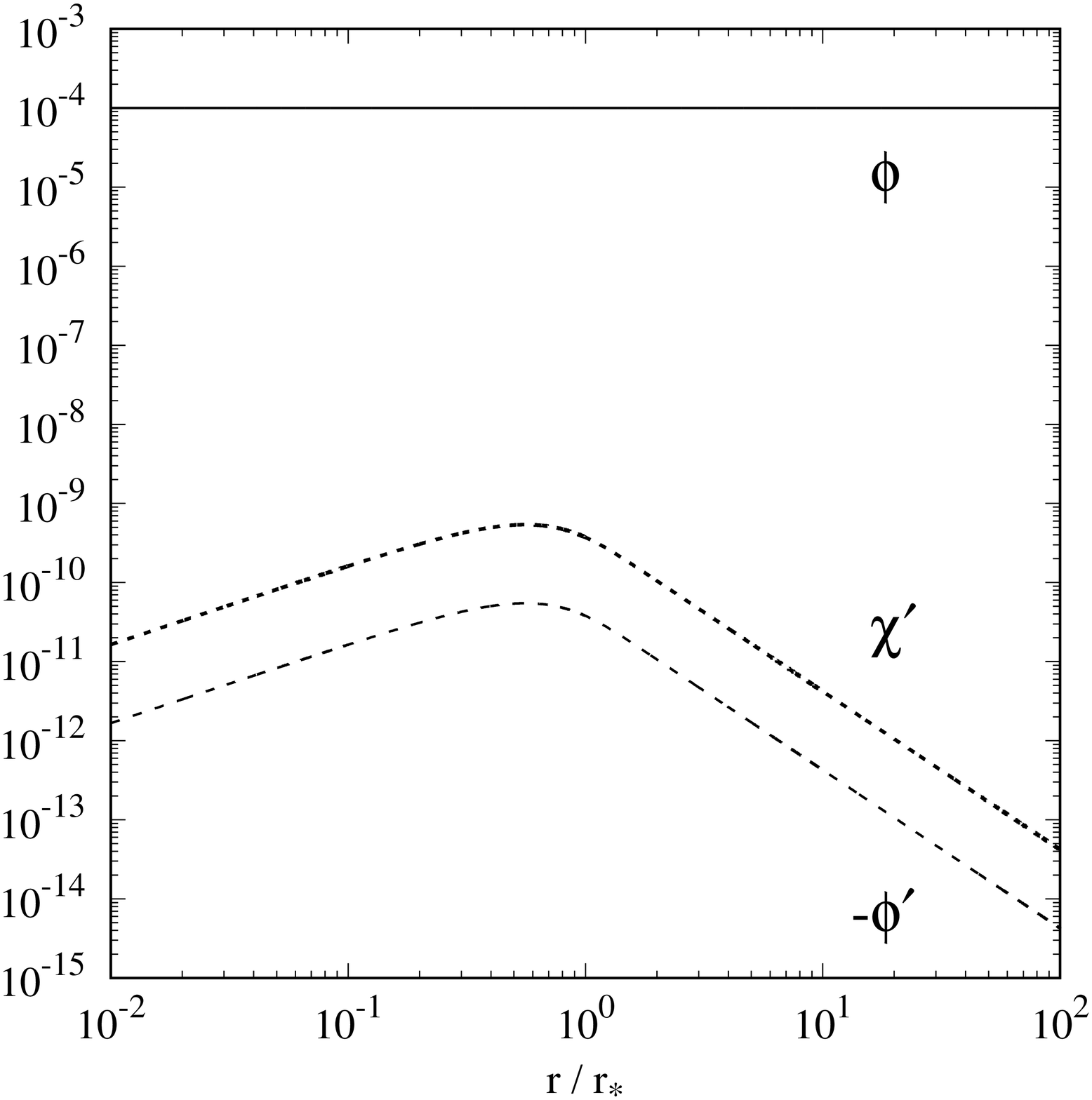}
\hspace{1cm}
\includegraphics[width=3.2in]{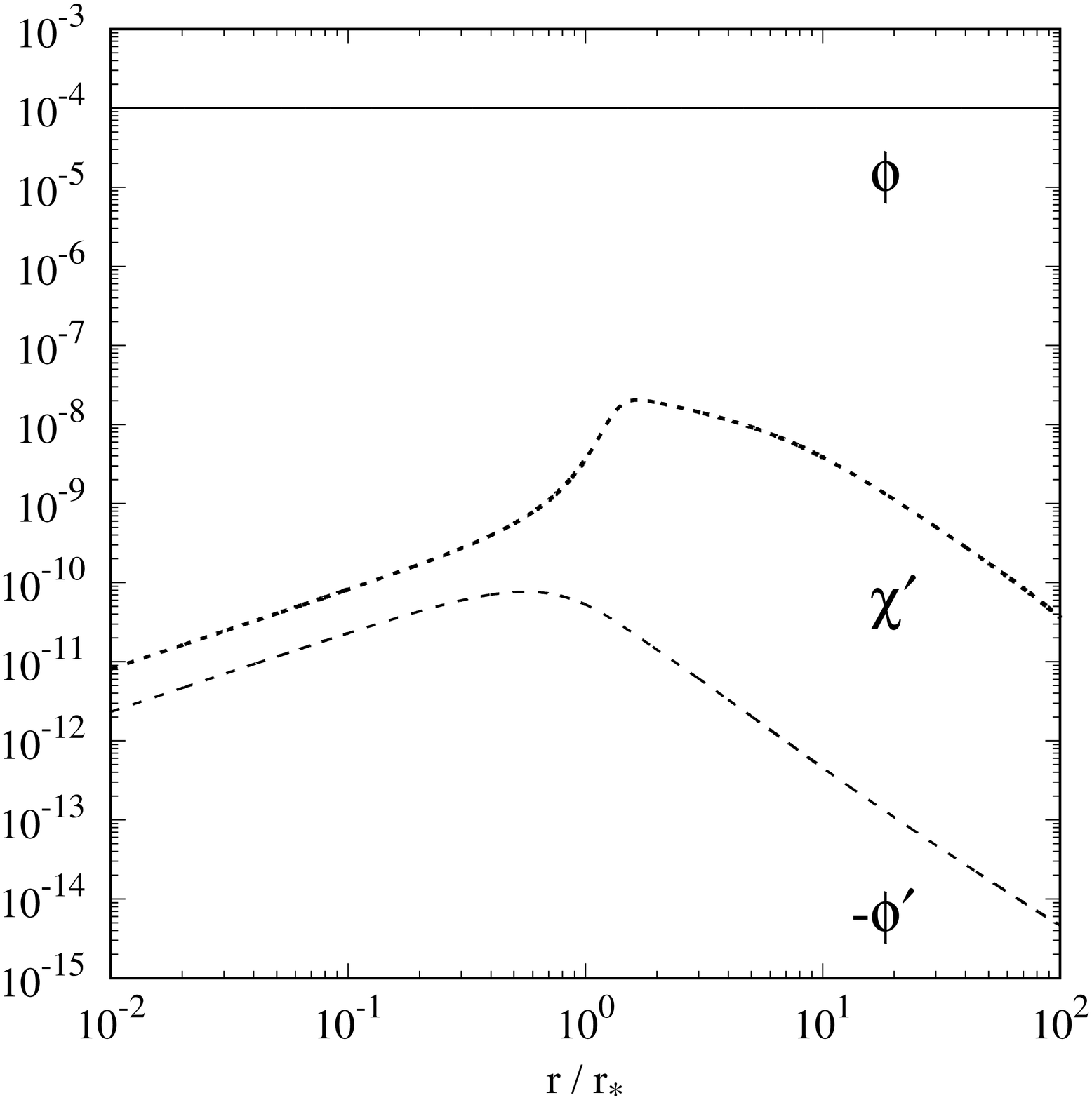}
\caption{Numerical solutions of $\phi(r)$, $-\phi'(r)$, and 
$\chi'(r)$ (normalized by $M_{\rm pl}$, $M_{\rm pl}/r_*$, 
and $M_{\rm pl}$, respectively) for $\Phi_*=10^{-5}$, 
$\phi_0=10^{-4}M_{\rm pl}$,  and 
the density profile (\ref{denprofile}) with $a=3$ 
and $\mu=10^{-24}$.
Each panel corresponds to the model parameters 
$s_{\beta_3}=10^{4}$, $Q=1.3$  (left) and 
$s_{\beta_3}=10^{-4}$, $Q=-0.3$ (right).
We choose the boundary conditions of $\phi(r)$ and
$\chi'(r)$ to be consistent with Eqs.~(\ref{phiG3}) 
and (\ref{chiG3}), respectively, at $r/r_*=10^{-3}$.
The boundary conditions of gravitational potentials 
are chosen to match with $\Psi_{\rm GR}$ and 
$\Phi_{\rm GR}$ given by Eq.~(\ref{gra1GR}).
\label{fig1}}
\end{center}
\end{figure}
%%%%%%%%%%%%%%%%%%%%%%%%%%%%

While we have derived analytic solutions to the vector field for 
the constant densities inside and outside the body, we 
will also study the case of the varying matter density given by 
\be
\rho(r)=\rho_0 e^{-ar^2/r_*^2}
+\mu \rho_0\,,
\label{denprofile}
\ee
where $a$ and $\mu$ are constants, with $a={\cal O}(1)$ 
and $\mu \ll 1$. 
The density is nearly constant for the distance $r \ll r_*$, 
but it starts to decrease rapidly around $r=r_*$ to approach 
the asymptotic value $\mu \rho_0$.
Numerically, we solve the full equations of motion (\ref{be4})--(\ref{be1}), (\ref{be3}), 
and (\ref{mattereq}) by using Eqs.~(\ref{phiG3}), (\ref{chiG3}),  (\ref{gra1GR}), 
and (\ref{pressure}) as the boundary conditions of $\phi, \chi', \Psi, \Phi, P$ 
around the center of the body ($r=10^{-3}r_*$).

We recall that the analytic vector-field profile was derived by 
employing the leading-order general-relativistic gravitational potentials. 
To check the consistency of this procedure, we also integrate 
Eqs.~(\ref{be1})--(\ref{be2}) by neglecting the contributions of 
the vector field and solve Eqs.~(\ref{phieq1})--(\ref{chieq1})
for $\phi(r)$ and $\chi'(r)$. 
We substitute the derived solutions of $\phi(r)$ and $\chi'(r)$ 
into Eqs.~(\ref{be1})--(\ref{be2}) and confirm that the corrections 
to $\Psi_{\rm GR}$ and $\Phi_{\rm GR}$ induced by 
the vector field remain small.
These approximate solutions exhibit good agreement 
with the full numerical results.

In the left panel of Fig.~\ref{fig1}, we plot the numerically integrated 
solutions to $\phi(r)$, $-\phi'(r)$, and 
$\chi'(r)$ versus $r/r_*$ for $s_{\beta_3}=10^{4}$, 
$Q=1.3$, $a=3$, $\mu=10^{-24}$, $\Phi_*=10^{-5}$, 
and $\phi_0=10^{-4}M_{\rm pl}$.
As estimated analytically from Eqs.~(\ref{phidr}) and (\ref{chiG3}), 
both $-\phi'(r)$ and $\chi'(r)$ are in proportion to $r$ for the 
distance $r \lesssim r_*$. For $r$ larger than $r_*$, they 
start to decrease according to Eqs.~(\ref{phiva}) and (\ref{chiva2}), i.e., 
$-\phi'(r) \propto 1/r^2$ and $\chi'(r) \propto 1/r^2$.
For all the distances of interest, the derivatives  
$|\chi'(r)|$ and $|\phi'(r)|$ are suppressed by the Vainshtein mechanism with 
$\phi$ nearly frozen around $\phi_0$. 

The right panel of Fig.~\ref{fig1} corresponds to the model parameters 
$s_{\beta_3}=10^{-4}$ and $Q=-0.3$.
Again, the solutions for the distance $r \lesssim r_*$ have the dependences 
$-\phi'(r) \propto r$ and $\chi'(r) \propto r$, but in this case there exists the 
intermediate regime ($r_*<r<r_c \approx 10r_*$) in which the longitudinal 
mode is given by Eq.~(\ref{chisr}).
For $r>r_c$, both $\chi'(r)$ and $-\phi'(r)$ decrease 
in proportion to $1/r^2$. 
Thus, even for $s_{\beta_3} \ll 1$, the Vainshtein mechanism 
leads to the suppression of both $|\chi'(r)|$ and $|\phi'(r)|$. 

We note that, for cubic scalar Galileons, the derivative of the longitudinal scalar $\chi(r)$ behaves as $\chi'(r) \propto r$ for 
$r<r_*$ and as $\chi'(r) \propto r^{-1/2}$ for $r_*<r<r_V$, 
where $r_V$ is the Vainshtein 
radius \cite{Burrage,Kase13,CKNT,APSS,BCLM}. 
Inside the body, the solution of $\chi'(r)$ derived above shares 
the common property to that of cubic scalar Galileons.
Outside the body, the solution found for the coupling $s_{\beta_3} \ll 1$ has a similar property [$\chi'(r) \propto r^{-1/2}$] to that for 
cubic scalar Galileons at the distance $r_*<r<r_c$. 
For the coupling $s_{\beta_3} \gg 1$, the exterior solution 
[$\chi'(r) \propto r^{-2}$]
decreases faster than that of cubic scalar Galileons.

\subsection{Gravitational potentials outside the body}

Let us estimate the corrections to leading-order gravitational potentials 
(\ref{gra2GR}) outside the body induced by the presence of the vector field coupled to matter. In doing so, we express 
Eqs.~(\ref{be1}) and (\ref{be2}) in the forms
\ba
&&
\frac{2M_{\rm pl}^2}{r}\Phi'
-\frac{M_{\rm pl}^2}{r^2}
\left( 1-e^{2\Phi} \right)
=e^{2\Phi} \rho+\Delta_{\Phi}\,,
\label{GR1d}\\
&&
\frac{2M_{\rm pl}^2}{r}\Psi'
+\frac{M_{\rm pl}^2}{r^2}
\left( 1-e^{2\Phi} \right)
=\Delta_{\Psi}\,,
\label{GR2d}
\ea
where $\Delta_{\Phi}$ and $\Delta_{\Psi}$ are
corrections to the gravitational equations in GR. 
Since we are interested in the behavior of $\Phi$ and 
$\Psi$ for $r>r_*$, we set $\rho=\mu \rho_0
=\mu \Phi_* M_{\rm pl}^2/r_*^2$ in Eq.~(\ref{GR1d}). 
For the integrations of Eqs.~(\ref{GR1d}) and (\ref{GR2d}), 
we employ the weak-gravity approximation (\ref{weakap}) 
in such a way that the general-relativistic corrections 
to gravitational potentials higher than second order
are neglected. In the following, we will consider the two different 
cases, (i) $s_{\beta_3} \gg 1$ and (ii)
$s_{\beta_3} \ll 1$, separately.

\subsubsection{$s_{\beta_3} \gg 1$}

In this case, we exploit the solutions from Eqs.
(\ref{phival})--(\ref{chiva2}) and substitute them into Eqs.~(\ref{be1})--(\ref{be2}). 
There exists the term of the form 
$\rho Q \phi_0^2\Phi_*/M_{\rm pl}^2$, which 
is at most of the order $\rho \Phi_*^2=
\mu M_{\rm pl}^2\Phi_*^3/r_*^2$ under 
the condition (\ref{Qeq}) by identifying $a_0$ 
with $\phi_0$. We neglect such contributions relative to 
the term of the order $\phi_0^2 \Phi_*^2/r_*^2$, 
which amounts to the condition 
$\mu \Phi_* \ll (\phi_0/M_{\rm pl})^2$.
Then, the corrections up to the order of $\Phi_*^2$ 
can be estimated as
\be
\Delta_{\Phi} \simeq
-\frac{(4\eta_1-1) \phi_0^2 r_*^2 \Phi_*^2}{72r^4}\,,
\qquad
\Delta_{\Psi} \simeq
-\frac{\phi_0^2 r_*^2 \Phi_*^2 }{72r^4}\,.
\label{gravicon2}
\ee
For the integration of Eqs.~(\ref{GR1d})--(\ref{GR2d}), 
we choose the integration constant in such a way that the solutions $\Phi=r_g/(2r)=r_*\Phi_*/(6r)$, 
$\Psi=-r_g/(2r)=-r_*\Phi_*/(6r)$ are recovered 
in the vacuum limit. 
Then, we obtain the integrated solutions corrected by 
the presence of the vector field, as
\ba
\Phi & \simeq & \frac{r_* \Phi_*}{6r} \left[ 
1+\frac{(4\eta_1-1)\Phi_*}{24} 
\left( \frac{\phi_0}{M_{\rm pl}} \right)^2 
\frac{r_*}{r} \right]\,,\label{Phi1}\\
\Psi & \simeq & -\frac{r_* \Phi_*}{6r} \left[ 
1+\frac{(2\eta_1-1)\Phi_*}{24} 
\left( \frac{\phi_0}{M_{\rm pl}} \right)^2 
\frac{r_*}{r} \right]\,.\label{Psi1}
\ea
The second terms in the square brackets of Eqs.~(\ref{Phi1}) and (\ref{Psi1}) induce the difference 
between the gravitational potentials, as
\be
\Phi+\Psi 
\simeq \beta_{\rm PN} U^2\,,
\ee
where
\be
\beta_{\rm PN} \equiv
\frac{\eta_1}{2}\left( \frac{\phi_0}{M_{\rm pl}} \right)^2\,,\qquad \quad U \equiv \frac{r_* \Phi_*}{6r}\,.
\ee
For $\phi_0 \lesssim M_{\rm pl}$, we have $\Phi+\Psi \lesssim U^2$.  
The quantity $\beta_{\rm PN}$
may be regarded as the second parametrized post-Newtonian 
parameter \cite{Willbook}. The experimental bound 
$|\beta_{\rm PN}| \lesssim 2.3 \times 10^{-4}$ 
from the Nordtvedt effect \cite{Will} can be satisfied for 
$\phi_0 \lesssim 10^{-2}M_{\rm pl}$. 
We note that the condition (\ref{Qa0}) translates to 
$\phi_0 \lesssim \sqrt{\Phi_*/|Q|}M_{\rm pl}$, so 
this can put a tighter limit of $\phi_0$ for 
$\Phi_*/|Q| \ll 10^{-4}$.

\subsubsection{$s_{\beta_3} \ll 1$}

For the coupling $s_{\beta_3}$ satisfying  
$s_{\beta_3} \ll (Q\mu)^2$, we employ the solutions from Eqs.
(\ref{phiG3sd}) and (\ref{chirsm}) for the estimations 
of $\Phi$ and $\Psi$. 
Up to the order of $\Phi_*^2$, the correction terms
$\Delta_{\Phi}$ and $\Delta_{\Psi}$ are given, 
respectively, by 
\be
\Delta_{\Phi} 
\simeq \frac{Q^2(1-\mu)^2 \phi_0^2 r_*^2 \Phi_*^2}{18r^4}\,,
\qquad
\Delta_{\Psi} 
\simeq -\frac{Q^2(1-\mu)^2 \phi_0^2 r_*^2 \Phi_*^2}{18r^4}\,.
\ee
The integrated solutions to Eqs.~(\ref{GR1d}) 
and (\ref{GR2d}) read
\ba
\Phi &\simeq& \frac{r_* \Phi_*}{6r} \left[ 
1-\frac{Q^2(1-\mu)^2\Phi_*}{6}
\left( \frac{\phi_0}{M_{\rm pl}} \right)^2 
\frac{r_*}{r} \right]\,,\label{Phi2}\\
\Psi &\simeq& -\frac{r_* \Phi_*}{6r} \left[ 
1-\frac{Q^2(1-\mu)^2\Phi_*}{6}
\left( \frac{\phi_0}{M_{\rm pl}} \right)^2 
\frac{r_*}{r}  \right]\,.\label{Psi2}
\ea
The second terms in Eqs.~(\ref{Phi2}) and (\ref{Psi2})
do not give rise to the difference between $\Phi$ and 
$-\Psi$, so $\beta_{\rm PN}=0$.

For the coupling satisfying $|Q|\mu \ll s_{\beta_3} \ll 1$, 
we substitute the solutions from Eqs. (\ref{phiG3sd}) and (\ref{chisr}) 
into Eqs.~(\ref{be1}) and (\ref{be2}). 
The correction terms $\Delta_{\Phi}$ and 
$\Delta_{\Psi}$ yield
\be
\Delta_{\Phi} \simeq
-\frac{\eta_2 \phi_0^2 \sqrt{s_{\beta_3}}\Phi_*}
{2\sqrt{r_*}r^{3/2}}\,,\qquad
\Delta_{\Psi} \simeq {\cal O} \left( \Phi_*^2 \right)\,.
\ee
Then, the integrated solutions to gravitational potentials 
are approximately given by 
\ba
\Phi &\simeq & \frac{r_* \Phi_*}{6r} \left[ 
1-\eta_2 \sqrt{s_{\beta_3}} 
\left( \frac{\phi_0}{M_{\rm pl}} \right)^2 
\left( \frac{r}{r_*} \right)^{3/2} \right]\,,
\label{Phi3}\\
\Psi &\simeq& -\frac{r_* \Phi_*}{6r} \left[ 
1+2\eta_2 \sqrt{s_{\beta_3}} 
\left( \frac{\phi_0}{M_{\rm pl}} \right)^2 
\left( \frac{r}{r_*} \right)^{3/2}\right]\,.
\label{Psi3}
\ea
The second terms in Eqs.~(\ref{Phi3}) and (\ref{Psi3}) lead 
to the difference between $\Phi$ and $-\Psi$, with the relative ratio 
\be
\gamma \equiv -\frac{\Phi}{\Psi} \simeq 
1-3\eta_2 \sqrt{s_{\beta_3}} 
\left( \frac{\phi_0}{M_{\rm pl}} \right)^2 
\left( \frac{r}{r_*} \right)^{3/2}\,.
\ee
At  $r=r_c=r_*/s_{\beta_3}^{1/3}$, 
the quantity $|\gamma-1|$ reaches the maximum 
value $|\gamma-1|_{\rm max}=3|\eta_2| (\phi_0/M_{\rm pl})^2$.
The local gravity bound $|\gamma-1|<2.3 \times 10^{-5}$ 
arising from the Cassini tracking \cite{Will} 
can be satisfied for $\phi_0 \lesssim 10^{-3}M_{\rm pl}$. 

\section{Quartic derivative couplings}
\label{quasec}

We proceed to the case of quartic derivative interactions given by 
\be
G_4=\frac{M_{\rm pl}^2}{2}+\beta_4 M_{\rm pl}^2 
\left( \frac{X}{M_{\rm pl}^2} \right)^n\,, 
\qquad G_{2,3,5,6}=0\,,
\label{G4model}
\ee
where $\beta_4$ is a dimensionless constant, and $n$ is a positive integer.
As we already mentioned in Sec.~\ref{eomsec}, we can express Eq.~(\ref{be5}) in the form (\ref{A1re}), where
\be
{\cal F}=2n \beta_4 \left( \frac{X}{M_{\rm pl}^2} \right)^{n-1}
\left[ f(1-h)+(f'-2nf')hr+2(n-1)h 
\frac{A_0 A_0'r-fX_1}{X} \right]\,, 
\ee
which vanishes in the limit that $\beta_4 \to 0$. 
Hence, the branch consistent with this limit corresponds to 
\be
\chi'=0\,.
\ee
Since there is no longitudinal propagation of the vector field, 
the temporal component $\phi$ alone can affect the solutions 
to gravitational potentials. 

Let us consider the matter density profile given by Eq.~(\ref{mudef}).
Inside the body, we substitute the leading-order gravitational 
potentials (\ref{gra1GR}) into Eq.~(\ref{be4}). 
Then, the temporal vector component obeys
\be
\frac{d}{dr} \left( r^2 \phi' \right) 
-\frac{\Phi_*}{\phi r_*^2} 
\left[ 4n \beta_4 M_{\rm pl}^2 \left( \frac{\phi^2}
{2M_{\rm pl}^2} \right)^n-(1+Q)\phi^2 
\right] r^2= 0\,.
\label{dphieqG4}
\ee
Assuming the solution in the form (\ref{phiasu}) 
and imposing the boundary condition $\phi'(0)=0$, the integrated solution 
to Eq.~(\ref{dphieqG4}) reads
\be
\phi(r)
\simeq \phi_0 \left[ 1-\frac{\Phi_*(1+Q-b_4)}{6} 
\frac{r^2}{r_*^2} \right]\,, 
\label{phiG4}
\ee
where 
\be
b_4\equiv 2^{2-n}n \beta_4 
\left( \frac{\phi_0}{M_{\rm pl}} \right)^{2(n-1)}\,.
\ee
Provided that $|b_4| \lesssim 1$, 
the second term in the square bracket of Eq.~(\ref{phiG4}) 
is at most of the order $\Phi_*$. 
We will assume this condition in the following discussion.

Outside the body, we substitute the leading-order external gravitational 
potentials (\ref{gra2GR}) into Eq.~(\ref{be4}). 
This leads to 
\be
\frac{d}{dr} \left( r^2 \phi' \right)
\simeq -\frac{Q \phi \Phi_*  \mu}{r_*^2}r^2\,.
\label{phioutG4}
\ee
The integrated solutions to $\phi'(r)$ and $\phi(r)$, which 
match with those in the regime $r<r_*$, are given by 
\ba
\phi'(r)
&\simeq&-\frac{\phi_0 \Phi_* r_*}{3r^2}
\left( {\cal H}+Q\mu \frac{r^3}{r_*^3}\right) \,,
\label{dphiG42} \\
\phi(r)
&\simeq& 
\phi_0 \left[ 1+\frac{\Phi_*}{6} 
\left\{ {\cal H} 
\left( \frac{2r_*}{r}-3 \right) 
-Q\mu \frac{r^2}{r_*^2} \right\} 
\right]\,,
\label{dphiG42d}
\ea
where 
\be
{\cal H} \equiv 1+Q(1-\mu) -b_4\,.
\ee
For the distance 
\be
r \ll \frac{r_*}{|Q \mu|^{1/3}}\,,
\label{rcon2}
\ee
the last terms on the rhs of 
Eqs.~(\ref{dphiG42}) and (\ref{dphiG42d}), which contain 
$Q\mu$, can be neglected.
The upper bound of Eq.~(\ref{rcon2}) is similar 
to the Vainshtein radius $r_V$ given by Eq.~(\ref{rV}). 
In this regime, the gravitational Eqs.~(\ref{be1}) 
and (\ref{be2}) are expressed in the forms (\ref{GR1d}) and 
(\ref{GR2d}), respectively, with the correction
terms (up to the order of $\Phi_*^2$)
\be
\Delta_{\Phi} 
\simeq 
\frac{\phi_0^2 r_*^2 \Phi_*^2 ({\cal H}-1)^2}{18r^4}\,,
\qquad
\Delta_{\Psi} 
\simeq 
\frac{b_4 \phi_0^2r_*\Phi_*(2{\cal H}-1)}{3r^3} 
-\frac{\phi_0^2 r_*^2 \Phi_*^2 ({\cal H}-1)^2}{18r^4}\,,
\label{delPsiG4}
\ee
where the terms of the order $b_4 \phi_0^2 r_*^2 
\Phi_*^2/r^4$ have been neglected relative to the first 
term in $\Delta_{\Psi}$.
Then, the integrated solutions to 
$\Phi$ and $\Psi$ are given, respectively, by 
\ba
\Phi &\simeq& \frac{r_* \Phi_*}{6r} \left[ 
1-\frac{\Phi_*({\cal H}-1)^2}{6} 
\left( \frac{\phi_0}{M_{\rm pl}} \right)^2 
\frac{r_*}{r} \right]\,,
\label{Phi4}\\
\Psi &\simeq& -\frac{r_* \Phi_*}{6r} \left[ 
1+b_4(2{\cal H}-1) \left( \frac{\phi_0}{M_{\rm pl}} \right)^2
-\frac{\Phi_*({\cal H}-1)^2}{6} 
\left( \frac{\phi_0}{M_{\rm pl}} \right)^2 
\frac{r_*}{r}
 \right]\,.\label{Psi4}
\ea
While the last terms of Eqs.~(\ref{Phi4}) and (\ref{Psi4}) do not give rise to 
the difference between $\Phi$ and $-\Psi$, the second term of Eq.~(\ref{Psi4}) 
leads to the difference with the relative ratio
\be
\gamma=-\frac{\Phi}{\Psi}
\simeq 1-b_4 (2{\cal H}-1) \left( \frac{\phi_0}{M_{\rm pl}} \right)^{2}\,.
\ee
Since the constant ${\cal H}$ is of the order unity, 
the experimental bound $|\gamma-1|<2.3 \times 10^{-5}$ 
can be satisfied under the condition 
\be
2^{2-n} n |\beta_4| \left( \frac{\phi_0}
{M_{\rm pl}} \right)^{2n} \lesssim 10^{-5}\,.
\label{boG4}
\ee
For $n=1$ and $|\beta_4|={\cal O}(1)$, 
the bound (\ref{boG4}) translates to 
$\phi_0 \lesssim 10^{-3}M_{\rm pl}$. 
As in the case of cubic vector Galileons, 
the coupled vector-field model 
with quartic derivative interactions is also 
consistent with local gravity constraints for 
$\phi_0$ much smaller than $M_{\rm pl}$.

\section{Quintic vector Galileons}
\label{quinsec}

We will study whether or not quintic derivative interactions 
give rise to solutions of the vector field operated by 
the Vainshtein mechanism. 
For concreteness, we consider the quintic vector 
Galileon given by the functions 
\be
G_5=\beta_5 \frac{X^2}{M_{\rm pl}^4}\,,\qquad 
G_4=\frac{M_{\rm pl}^2}{2}, \qquad 
G_{2,3,6}=0\,,
\label{G5lag}
\ee
where $\beta_5$ is a dimensionless constant. 
We choose the matter density profile (\ref{mudef}), 
but we will also discuss the case in which the density 
varies outside the body.

\subsection{$r<r_*$}

Inside the spherically symmetric body, 
we substitute the leading-order gravitational 
potentials of Eq. (\ref{gra1GR}) into Eqs.~(\ref{be4})--(\ref{be5}) 
and then expand them up to the first order of $\Phi_*$. 
Then, it follows that 
\ba
& &
\frac{d}{dr} \left( r^2 \phi' \right)+(1+Q)\phi \Phi_* 
\frac{r^2}{r_*^2}+\frac{\beta_5 \phi}{3M_{\rm pl}^4 
r_*^2} \left[ 6r_*^2 \chi'^2 \chi''+r \Phi_* 
(r \phi^2 \chi''+2\phi^2 \chi'-3\chi'^3) \right] \simeq 0\,,
\label{be4G5}\\
& &
3M_{\rm pl}^4 Q\Phi_* r^2 \chi'+\beta_5 \left[ 
\phi \phi' \left( \Phi_*r^2 \phi^2+6r_*^2 \chi'^2 \right)+
2\Phi_*r \chi'^4 \right] \simeq 0\,.
\label{be5G5}
\ea
As in the case of cubic vector Galileons, we search 
for analytic solutions to $\phi'(r)$ and $\chi'(r)$ 
proportional to $r$ around the center of the body. 
Provided that $\phi$ is nearly constant around $\phi_0$, 
the first three terms on the lhs of Eq.~(\ref{be5G5}) 
are in proportion to $r^3$, whereas the last term has 
the dependence $2\beta_5\Phi_*r\chi'^4 \propto r^5$.
Neglecting the last term of Eq.~(\ref{be5G5}) 
around $r=0$, we can solve Eq.~(\ref{be5G5}) for $\chi'$ as 
\be
\chi'(r) \simeq -\frac{M_{\rm pl}^4 Q\Phi_*r^2}
{4\beta_5 \phi \phi' r_*^2} \left[ 
1-\sqrt{1-\frac{8\beta_5^2 \phi^4 \phi'^2r_*^2}
{3M_{\rm pl}^8Q^2\Phi_* r^2}} \right]\,, 
\label{chiG5}
\ee
where we have chosen the branch recovering $\chi' \to 0$ 
for $\beta_5 \to 0$. 
Provided that the longitudinal mode $\chi'$ is suppressed 
relative to $\phi$, the $\beta_5$-dependent terms 
in Eq.~(\ref{be4G5}) should work as corrections to 
the leading-order solution 
\be
\phi_{\rm leading}(r)=\phi_0 \left[ 1 
-\frac{(1+Q)\Phi_*}{6} \frac{r^2}{r_*^2} \right]\,,
\label{phile}
\ee
which is nearly constant around $\phi_0$.
Substituting the derivative of Eq.~(\ref{phile}) into 
Eq.~(\ref{chiG5}), the leading-order longitudinal 
mode reads
\be
\chi'_{\rm leading}(r) \simeq \frac{3M_{\rm pl}^4 Qr}
{4\beta_5 (1+Q)\phi_0^2} \left( 1-\sqrt{1-\varepsilon_5} 
\right)\,, \qquad \quad 
\varepsilon_5 \equiv 
\frac{8(1+Q)^2 \beta_5^2 \Phi_* \phi_0^6}
{27Q^2 M_{\rm pl}^8r_*^2}\,.
\label{chi2G5}
\ee
For the existence of this solution, we require the condition 
$\varepsilon_5 \le 1$, which translates to
\be
\beta_5^2 \le \frac{27Q^2}{8(1+Q)^2} 
\frac{M_{\rm pl}^8r_*^2}{\Phi_* \phi_0^6}\,,
\ee
so that  the coupling $\beta_5$ is bounded from above.

Let us consider the case in which the condition 
\be
\varepsilon_5 \ll 1
\label{epcon5}
\ee
is satisfied. Assuming that both $\beta_5$ and $Q$ are 
positive, the longitudinal mode (\ref{chi2G5}) reduces to 
\be
\chi'_{\rm leading}(r) \simeq \frac{\phi_0}{12} 
\sqrt{6\Phi_* \varepsilon_5}\frac{r}{r_*}\,, 
\label{chile}
\ee
which means that $\chi'^2_{\rm leading}(r)$ is suppressed relative to $\phi_0^2$ by the factor $(\Phi_* \varepsilon_5/24)(r/r_*)^2$.
By using the solution (\ref{chile}), the $\beta_5$-dependent 
terms in Eq.~(\ref{be4G5}) can be estimated as 
$3Q\phi_0 \varepsilon_5 \Phi_* r^2/[(1+Q)r_*^2]$. 
Taking into account this contribution and integrating Eq.~(\ref{be4G5})  with respect to $r$, we obtain
\be
\phi(r) \simeq \phi_0 \left[ 1-\frac{\{8(1+Q)^2+3Q\varepsilon_5\}
\Phi_*}{48(1+Q)} 
\frac{r^2}{r_*^2} \right]\,,
\ee
which is close to the leading-order temporal 
component (\ref{phile}).

The solution (\ref{chile}) corresponds to the case in which the first two terms of Eq.~(\ref{be5G5}) balance each other. 
On using Eq.~(\ref{phile}), the third and fourth terms of Eq.~(\ref{be5G5}) are suppressed relative to 
the first term by the factors $-\varepsilon_5/4$ 
and $\varepsilon_5^2 \Phi_*r^2/[96(1+Q)r_*^2]$, respectively.
Then, the longitudinal vector component yields
\be
\chi'(r) \simeq \frac{\phi_0}{12} 
\sqrt{6\Phi_* \varepsilon_5}\frac{r}{r_*}
\left[ 1+\frac{\varepsilon_5}{4} \left\{ 
1-\frac{\varepsilon_5\Phi_*}{24(1+Q)} 
\frac{r^2}{r_*^2} \right\} \right]\,,
\label{chile3}
\ee
and hence $\chi'(r)$ is well described by the leading-order 
solution (\ref{chile}) for $\varepsilon_5 \ll 1$.

\subsection{$r>r_*$}

Substituting the leading-order gravitational potentials 
of Eq. (\ref{gra2GR}) into Eqs.~(\ref{be4})--(\ref{be5}) and 
expanding them up to the first order of $\Phi_*$, the field 
equations outside the body yield
\ba
& &
\frac{d}{dr} \left( r^2 \phi' \right)+Q\mu \phi \Phi_* 
\frac{r^2}{r_*^2}+\frac{\beta_5 \phi}{3M_{\rm pl}^4 
r^2} \left[ 6r^2 \chi'^2 \chi''+r_* \Phi_* 
(r \phi^2 \chi''-\phi^2 \chi'+3\chi'^3) \right] \simeq 0\,,
\label{be4G5out}\\
& &
3M_{\rm pl}^4 Q\mu \Phi_* r^4 \chi'+\beta_5r_*^2 \left[ 
\phi \phi' r \left( \Phi_* r_* \phi^2+6r \chi'^2 \right)+
2\Phi_* r_* \chi'^4 \right] \simeq 0\,.
\label{be5out}
\ea
Provided that the last term of Eq.~(\ref{be5out}) is much  
smaller than the other terms, Eq.~(\ref{be5out}) can 
be explicitly solved as
\be
\chi'(r) \simeq -\frac{M_{\rm pl}^4 Q\mu \Phi_*r^2}
{4\beta_5 \phi \phi' r_*^2} \left[ 
1-\sqrt{1-\frac{8\beta_5^2 \phi^4 \phi'^2r_*^5}
{3M_{\rm pl}^8Q^2 \mu^2 \Phi_* r^5}} \right]\,.
\label{chiG52}
\ee
In the vacuum limit ($\mu \to 0$), the solution to 
Eq.~(\ref{be4G5out}) derived by neglecting the contributions 
of $\chi'$ and $\chi''$ reads $\phi'(r) \propto 1/r^2$ 
and $\phi(r) \simeq {\rm constant}$. 
Then, the second term in the square root 
of Eq.~(\ref{chiG52}), which is always positive, 
exhibits the divergence for $\mu \to 0$. 
If we consider a rapidly decreasing 
density profile like Eq.~(\ref{denprofile}), 
the solution to Eq.~(\ref{chiG52}) becomes imaginary 
above the distance $r_d$ at which the second term in the 
square root of Eq.~(\ref{chiG52}) is equivalent to 1. 
Unless we consider slowly varying 
density profiles like $\rho(r)=\rho_0(r_*/r)^p$ with 
$p<9/2$, it is not possible to realize the existence of 
regular solutions for the distance $r>r_d$.

The solution in Eq. (\ref{chiG52}) has been derived by neglecting the 
contribution of the last term of Eq.~(\ref{be5out}).
As we estimated in Eq.~(\ref{chile3}), this term works 
as a tiny correction to the leading-order longitudinal mode 
(\ref{chi2G5}) around $r=r_*$. 
For the density profile (\ref{denprofile}), 
we numerically integrate the vector-field equations of motion  
coupled to the gravitational equations and find that the last 
term of Eq.~(\ref{be5out}) remains small relative to 
the other terms for the distance $r_*<r<r_{d}$.
For $r>r_d$, there are no real solutions to the longitudinal mode.
Thus, the quintic vector Galileon does not allow the 
existence of consistent solutions to $\chi'(r)$ for 
realistic density profiles that rapidly decrease for $r>r_*$.

We also studied the model of the linear coupling 
$G_5(X)=\beta_5 X/M_{\rm pl}^2$ and found that 
the solution to $\chi'$ has a similar property to 
that of the model (\ref{G5lag}). 
In this case, the $\chi'^4$ term is absent unlike Eq.~(\ref{be5out}), 
so we have the exact solution to $\chi'(r)$ analogous to  
Eq.~(\ref{chiG52}).
Hence, it is not possible to realize the solution of $\chi'(r)$ 
regular throughout the region $r>r_*$ for 
the rapidly decreasing matter density. 
Thus, the models with quintic derivative couplings 
are generally plagued by the absence of regular 
external solutions operated by the Vainshtein mechanism.

\section{Sixth-order derivative couplings}
\label{sixthsec}

Finally, we study the model with sixth-order derivative 
interactions given by 
\be
G_{6} = \frac{\beta_{6}}{M_{\rm pl}^{2}} \left(\frac{X}{M_{\rm pl}^{2}}\right)^{n} \,, \qquad
G_{4} = \frac{M_{\rm pl}^{2}}{2} \,,\qquad
G_{2,3,5} = 0 \,, 
\label{G6model}
\ee
where $\beta_{6}$ is a dimensionless constant, and $n$ is 
a positive integer of order 1.
According to the discussion given in Sec.~\ref{eomsec}, 
the branch consistent with the limit 
$\beta_{6} \to 0$ corresponds to 
\be
\chi'=0\,.
\ee

Let us consider the matter density profile (\ref{mudef}).
Inside the body, we substitute Eq.~(\ref{gra1GR}) into Eq.~(\ref{be4}) and expand it up to second order 
in $\Phi_*$ for the terms containing $\beta_6$. 
Then, the temporal vector component obeys
\be
\frac{d}{dr}\left( r^{2} \phi' \right)+\phi \Phi_*
 \frac{r^2}{r_*^2} 
\left[ 1+Q -\frac{2b_6 ( Q\Phi_* \phi^2
-n\phi'^2 r_*^2)}{(1+2b_6 \Phi_*)\phi^2} \right]
\simeq 0\,,
\label{sixtheom}
\ee
where
\be
b_6 \equiv \frac{\beta_6}{3M_{\rm pl}^2 r_*^2} 
\left( \frac{\phi^2}{2M_{\rm pl}^2} \right)^n\,.
\ee
In the limit that $b_6 \to 0$, we obtain the leading-order 
solution $\phi_{\rm leading}(r)=
\phi_0 [1-\Phi_*(1+Q)r^2/(6r_*^2)]$, so 
$\phi(r)$ is nearly frozen around $\phi_0$. 
Assuming that $|b_6 \Phi_*|\ll 1$, the terms 
containing $b_6$ in Eq.~(\ref{sixtheom}) can be 
regarded as the corrections to $\phi_{\rm leading}(r)$.
Ignoring the terms higher than $\Phi_*^2$ 
inside the square bracket of Eq.~(\ref{sixtheom}) and 
dealing with $b_6$ as a constant with 
$\phi (r) \simeq \phi_0$, 
the resulting field derivative is given by 
\be
\phi'(r) \simeq -\frac{\phi_0 \Phi_*}{3r_*^2} 
\left( 1+Q-2Q b_6 \Phi_* \right)r\,, 
\label{phidG6}
\ee
where we have dropped the terms of the order $\Phi_*^2$ 
which are not multiplied by $b_6$. 
Since $|b_6 \Phi_*|\ll 1$, the correction induced by 
the coupling $b_6$ to the leading-order solution 
of $\phi'(r)$ is negligibly small.

Outside the body, we substitute 
Eq.~(\ref{gra2GR}) into Eq.~(\ref{be4}), 
solve it for $\phi''(r)$, and then 
expand it up to second order in $\Phi_*$ for the terms 
containing $b_6$.
This process leads to 
\be
\frac{d}{dr}(r^{2} \phi') \simeq -Q \mu \phi \Phi_{\ast} \frac{r^{2}}{r_{\ast}^{2}}
+\frac{2b_6 \Phi_* r_*^3}{\phi r^4} 
\left[ Q \mu \phi^2 \Phi_{\ast} \frac{r^{3}}{r_{\ast}^{2}}
+\Phi_* r_* \phi^2+\frac13 \phi' 
\left( 9\phi+n^2 r_* \Phi_* \phi' \right)r^2
-nr^3 \phi'^2 \right]\,.
\label{phioutG6}
\ee
Provided that the terms multiplied by $b_6$ are 
negligibly small relative to the term 
$-Q\mu \phi \Phi_* r^2/r_*^2$, 
the integrated solution for $r>r_*$, which matches Eq.~(\ref{phidG6}) at $r=r_*$, reads
\be
\phi'_{\rm leading}(r)= -\frac{\phi_0 \Phi_* r_*}{3r^2}
\left[ 1+Q(1-\mu) 
+Q\mu \frac{r^3}{r_*^3}\right] \,,
\label{dphioutG6}
\ee
where we have dropped the contribution 
$-2Q b_6 \Phi_*$ in Eq.~(\ref{phidG6}). 
The first term in the square bracket of Eq.~(\ref{phioutG6}) 
is suppressed by the factor $b_6 \Phi_*r_*^3/r^3$ 
relative to the first term on the rhs of Eq.~(\ref{phioutG6}). 
For the distance $r \ll r_*/|Q \mu|^{1/3}$, the leading-order field derivative is given by 
$\phi_{\rm leading}'(r) \simeq -\phi_0\Phi_*r_*
[1+Q(1-\mu)]/(3r^2)$, so the $b_6$-dependent terms 
in Eq.~(\ref{phioutG6}) can be estimated as
$-2\phi_0 Q(1-\mu)b_6 \Phi_*^2 r_*^4/r^4$. 
Then, the correction to $\phi'_{\rm leading}(r)$, 
which arises from the term containing 
$b_6$,  yields 
\be
\Delta \phi'(r)=\frac{2Q(1-\mu)\phi_0b_6 \Phi_*^2 
r_*^4}{3r^5}\,,
\ee
which is suppressed by the factor $Qb_6 \Phi_* r_*^3/r^3$ 
compared to Eq.~(\ref{dphioutG6}).

On using the leading-order solution (\ref{dphioutG6}) 
outside the body,  the correction terms in the gravitational Eqs.~(\ref{be1}) and (\ref{be2}), expanded up to 
the order of $\Phi_*^2$, are given, respectively,  by
\be
\Delta_{\Phi} 
\simeq \frac{Q^{2} (1-\mu )^{2} \phi_{0}^{2} \Phi_{\ast}^{2}r_{\ast}^{2}}{18 r^{4}} \,,\qquad
\Delta_{\Psi}
\simeq  
-\frac{Q^{2}(1-\mu )^{2} \phi_{0}^{2}r_*^2 
\Phi_*^{2}(r^2-12b_6r_*^2)}{18r^6}\,.
\label{deltapsiG6}
\ee
Then, the gravitational potentials 
induced by the vector field can be estimated as
\ba
\Phi &\simeq& \frac{r_{\ast} \Phi_{\ast}}{6 r} 
\left[ 1 - \frac{\Phi_{\ast} Q^{2}(\mu - 1)^{2}}{6} \left( \frac{\phi_{0}}{M_{\rm pl}}\right)^{2} 
\frac{r_{\ast}}{r}
\right] \,,
\\
\Psi &\simeq& -\frac{r_{\ast} \Phi_{\ast}}{6 r} 
\left[ 1 -\frac{\Phi_{\ast} Q^{2}(\mu - 1)^{2}}{6} \left( \frac{\phi_{0}}{M_{\rm pl}}\right)^{2} 
\frac{r_{\ast}}{r} \left( 1-3b_6 \frac{r_*^2}{r^2} 
\right) \right] \,.
\ea
The coupling $b_6$ induces the difference between $\Phi$ 
and $-\Psi$, such that 
\be
\Phi+\Psi \simeq \beta_{\rm PN}U^2\,,\qquad \quad
\beta_{\rm PN}=-3Q^2(1-\mu)^2b_6 
\left( \frac{\phi_0}{M_{\rm pl}} \right)^2 
\frac{r_*^2}{r^2}\,.
\ee
The parameter $\beta_{\rm PN}$ has a maximum 
at $r=r_*$ and decreases for larger $r$. 
The bound $|\beta_{\rm PN}|<2.3 \times 10^{-4}$ is well 
satisfied for $|Q| \sqrt{|b_6|}\,\phi_0/M_{\rm pl} 
\lesssim 10^{-2}$.

\section{Conclusions}
\label{consec}

We studied the propagation of the vector field coupled to 
matter in the form (\ref{Lcoupling}) in generalized Proca theories with derivative self-interactions. The difference from the previous analysis \cite{screening} is the existence of matter-vector couplings 
in the form (\ref{Lcoupling}) 
with all the derivative interactions taken into account up to sixth order. On the spherically symmetric and static background, there exists a temporal 
vector component $A_0$ besides a longitudinal scalar $\chi$.
To extract the constant mode $a_0$ in $A_0$ from the matter continuity 
equation, we defined the density $\rho$ and the 
pressure $P$ in the forms (\ref{rhodef}) and (\ref{Ppre}), respectively.
The matter-coupling term induced by the constant mode of $A_0$ is smaller than 
the intrinsic pressure under the condition (\ref{Qconas}), which translates 
to the inequality (\ref{Qa0}).

In Sec.~\ref{cubsec}, we derived the analytic vector-field profile both inside 
and outside a spherically symmetric body in the presence of cubic Galileon 
interactions by employing the general-relativistic gravitational potentials 
(\ref{gra1GR}) and (\ref{gra2GR}). 
This procedure can be justified under the conditions of Eq. (\ref{Qcon}), whose consistency was checked after deriving solutions to the vector field. We showed that both the longitudinal and temporal vector components are sufficiently suppressed 
due to the operation of the Vainshtein mechanism. 
This result was also numerically confirmed for the decreasing density 
profile (\ref{denprofile}). 
We computed corrections to the general-relativistic gravitational 
potentials outside the body and found that the model can be consistent 
with local gravity constraints even for the coupling $Q$ of order unity.

For the derivative couplings $G_i(X)$ with even indices $i$, the solution
to the longitudinal mode, which is consistent with the continuous limit of 
small couplings, corresponds to $\chi'=0$.
In Sec.~\ref{quasec}, we obtained solutions to the temporal vector component 
and the gravitational potentials for quartic power-law couplings given 
by Eq.~(\ref{G4model}). This model is compatible with Solar-System
constraints under the bound (\ref{boG4}). 
In Sec.~\ref{sixthsec}, we also carried out the similar analysis for  
sixth-order power-law couplings (\ref{G6model}) and showed the 
compatibility of solutions with local gravity tests.

In Sec.~\ref{quinsec}, we studied the vector-field profile in the presence
of quintic Galileon interactions.
We found that there are no consistent solutions of the longitudinal 
mode in the vacuum limit outside the body. 
This fact does not allow the existence of regular solutions for a realistic compact object whose density rapidly decreases outside the body. 
Thus, the quintic vector Galileon is the special case in 
which the screening mechanism for the longitudinal mode 
does not work. It remains to be seen whether this property 
also holds for general quintic derivative interactions
with matter couplings other than Eq.~(\ref{Lcoupling}).

While we focused on the behavior of the vector field on the spherically 
symmetric and static background within the Vainshtein 
radius, it will be of interest to explore how the effect of the 
matter-vector coupling leads to the 
modification to the cosmological dynamics in uncoupled generalized 
Proca theories studied in Ref.~\cite{cosmo}.
In particular, the signatures of weak gravity found in Refs.~\cite{Geff,NKT} 
for scales relevant to the growth of large-scale structures may be 
compensated by the matter-coupling term. 
This may provide us with the possibility of distinguishing between 
the coupled and uncoupled dark energy models 
constructed in generalized Proca theories.
This issue is left for a future work.

%%%%%%%%%%%%%%%%%%%%%%%%
\section*{Appendix: Coefficients in the gravitational 
equations}
%%%%%%%%%%%%%%%%%%%%%%%%

In Eqs.~(\ref{be1})--(\ref{be3}) the coefficients 
$c_{1,2,\dots,19}$ are given by 
\ba
&&c_1=-A_1 X G_{3,X}\,,\quad 
c_2=-2 G_4+4 ( X_0+2 X_1 ) G_{4,X}+8 X_1 X G_{4,XX}\,,\notag\\
&&c_3=-A_1 (3 hX_0+5h X_1-X) G_{5,X}
-2h A_1X_1 X G_{5,XX}\,,\notag\\
&&c_4=G_2-2 X_0 G_{2,X}-\frac{h}{f} \left(A_0A_1 A_0'+2f X  A_1' \right) G_{3,X}
+ \frac{h {A_0'}^2}{2 f} (2 g_4 - 1+ \frac{2 A_0^2}{f} g_{4,X})\,,\notag\\
&&c_5=-4 hA_1X_0G_{3,X}-4  h^2A_1 A_1' G_{4,X}
+\frac{8h}{f} (A_0X_1 A_0'- fh A_1 X A_1' ) G_{4,XX}+
 {\frac {2 h^2A_1 {A_0'}^2 (g_5+2 X_0 g_{5,X})}{f}}\,,\notag\\
&&c_6= 2( 1- h ) G_4+4 ( hX- X_0) G_{4,X}+8 h X_0X_1G_{4,XX}
-\frac{h}{f}\left[ (h-1) A_0A_1 A_0'+2 f (3hX_1+hX_0-X)  A_1' \right] G_{5,X} \notag\\
&&\hspace{.75cm}
-{\frac {2 h^2X_1}{f}} ( A_0A_1 A_0'+2f X A_1' ) G_{5,XX}
 +\frac{h{ A_0'}^2}{f} \left[ ( h-1 ) G_6+2 ( hX-X_0) G_{6,X}
 +4 hX_0X_1G_{6,XX}\right]\,,\notag\\
&&c_{7}=-G_2+2 X_1 G_{2,X}-\frac {hA_0A_1 A_0'G_{3,X}}{f}
-\frac {h{ A_0'}^2}{2f}( 2g_4 - 1 - 2 h {A_1}^{2} g_{4,X})\,,\notag\\
&&c_{8}=4 hA_1X_1G_{3,X}
+\frac{4hA_0 A_0' ( G_{4,X}+2 X_1G_{4,XX} )}{f}
-\frac{ 2h^2A_1 { A_0'}^2 (3 g_5+2 X_1g_{5,X})}{f}\,,\notag\\
&&c_{9}= 2( h-1 ) G_4-4 ( 2 h-1 )X_1 G_{4,X}-8 hX_1^2G_{4,XX}
-\frac{hA_0A_1A_0'}{f} [ (3 h-1) G_{5,X}+2hX_1G_{5,XX} ]\notag\\
&&\hspace{.75cm}
-\frac{h A_0'^2}{f} \left[ ( 3 h-1 ) G_6+2 (6 h-1)X_1G_{6,X}+4 hX_1^2G_{6,XX} \right]\,,\notag \\
&&c_{10}=-\frac {2h ( G_4-2 X G_{4,X} ) }{f}\,,\quad
c_{11}=-\frac { 2h^2A_1 X G_{5,X}}{f}\,,\quad 
c_{12}=\frac {h [ G_4-2(2 X_0+X_1) G_{4,X}-4 X_0 X G_{4,XX} ] }{f^2}\,,\notag\\
&&c_{13}={\frac { h^2 A_1 [ ( 3 X_0+X_1) G_{5,X}+2X_0 X G_{5,XX}] }{f^2}}\,,\quad
c_{14}=-{\frac {h A_1 [( 3 X_0+5 X_1) G_{5,X}+2 X_1 X G_{5,XX} ] }{f}}\,,\notag\\
&&c_{15}=\frac{h}{f^2}\big[
2 fA_1X_0G_{3,X}+2( 2 A_0 A_0'-fhA_1A_1' ) G_{4,X}
+4 \{ A_0 ( 2 X_0+X_1 )  A_0'-fhA_1 X A_1' \} G_{4,XX}\notag\\
&&\hspace{.75cm}-hA_1 A_0'^2( g_5+2X_0 g_{5,X} )  
\big]\,,\notag\\
&&c_{16}=-\frac{h}{f^2} \big[ 2f (G_4-2 X G_{4,X}+4 X_0X_1G_{4,XX})
+ h\{ 3 A_0A_1 A_0'+ 2f(X_0+3 X_1 )A_1' \} G_{5,X}\notag\\
&&\hspace{.95cm}+2h \{ A_0A_1 ( X_1+2 X_0 ) A_0'+2f X_1 X A_1' \} G_{5,XX}
+h{ A_0'}^2( G_6+2 X G_{6,X}+4X_0X_1G_{6,XX} ) \big]\,,\notag \\
&&c_{17}=-2 G_4+8 X_1 ( G_{4,X}+X_1G_{4,XX})\notag\\
&&\hspace{.95cm}+\frac{hA_0'}{f}\big[A_0A_1 ( 3 G_{5,X}+2 X_1G_{5,XX} )
+{ A_0'}\{ 3 G_6+4X_1(3 G_{6,X}+X_1G_{6,XX}) \} \big]\,,\notag\\
&&c_{18}=2 G_2
-\frac{2h}{f}\big[( A_0A_1 A_0'+2f X_1 A_1' ) G_{3,X}
+2 ( A_0A_0''+A_0'^2 ) G_{4,X}
+ 2 A_0' ( 2 X_0 A_0' -hA_0A_1 A_1' ) G_{4,XX}\big]\notag\\
&&\hspace{.95cm}+\frac{2h^2A_0'}{f^2} \big[ f  ( 2 A_1A_0''+ A_0'  A_1' ) g_5
+A_0'  ( A_0A_1 A_0'+2f X_1 A_1') g_{5,X}\big]
- \frac{h {A_0'}^2}{f} (2 g_4 - 1) \,,\notag\\
&&c_{19}=\frac{2 h}{f} \Big[  -2( A_0 A_0'+ fhA_1A_1' ) G_{4,X}
+4 X_1 ( A_0 A_0'-fhA_1A_1' ) G_{4,XX}
+h ( A_1A_0'^2+A_0 A_0' A_1'+ A_0A_1A_0'' ) G_{5,X}\notag\\
&&\hspace{.95cm}+2 hA_0'  ( A_0 X_1 A_1' + A_1X_0A_0') G_{5,XX}
+2 h A_0'A_0'' G_6
+\frac{hA_0'}{f}\Big\{( A_0A_0'^2+4 fX_1A_0'' -3 fh A_1 A_0' A_1' ) G_{6,X}\notag\\
&&\hspace{.95cm}+2 A_0'X_1 ( A_0 A_0'-fhA_1 A_1') G_{6,XX}\Big\} \Big]\,.
\ea
\section*{Acknowledgements}

R.~K. is supported by the Grant-in-Aid for Young Scientists B of the JSPS No.\,17K14297. 
S.~T. is supported by the Grant-in-Aid for Scientific Research Fund of the JSPS No.~16K05359 and 
the MEXT KAKENHI Grant-in-Aid for 
Scientific Research on Innovative Areas ``Cosmic Acceleration'' (No.\,15H05890).

%%%%%%%%%%%%%%%%

\end{document}